\pdfoutput=1
\documentclass[12pt,prd,preprint,eqsecnum,nofootinbib,amsmath,amssymb,
               tightenlines]{revtex4}

\usepackage{color,latexsym,graphics,graphicx,epsfig}
\usepackage{graphicx,graphics,color,latexsym}% Include figure files
\usepackage{hyperref}
\usepackage{ifthen}
\newboolean{codedoc}
\setboolean{codedoc}{true}
\usepackage{graphicx}% Include figure files
\usepackage{bm}% bold math
\def\dd{{\rm d}}

\def\x{{\bm x}}

\def\p{{\bm p}} 
\def\k{{\bm k}}

\def\st{\begin{equation}}
\def\stp{\end{equation}}
\def\bg{\begin{eqnarray}}
\def\nd{\end{eqnarray}}
\def\Eq#1{Eq.~(\ref{#1})}
\def\App#1{Appendix~\ref{#1}}
\def\Fig#1{Fig.~\ref{#1}}
\def\Sect#1{Section~\ref{#1}}

\def\pr{\mathcal P}
\def\drangle{\rangle\!\rangle}
\def\dlangle{\langle\!\langle}
\def\aveeps{\dlangle \epsilon_2 \drangle}

\def\Dlangle{\left\langle\!\!\left\langle}
\def\Drangle{\right\rangle\!\!\right\rangle}
\def\llangle{\left\langle}
\def\rrangle{\right\rangle}

\def\Eq#1{Eq.~(\ref{#1})}

\def\Fig#1{Fig.~\ref{#1}}
\def\Sect#1{Section~\ref{#1}}
\def\Ref#1{Ref.~\cite{#1}}

\def\nott#1{\setbox0=\hbox{$#1$}                % set a box for #1
   \dimen0=\wd0                                % and get its size
   \setbox1=\hbox{/} \dimen1=\wd1               % get size of /
   \ifdim\dimen0>\dimen1                        % #1 is bigger
      \rlap{\hbox to \dimen0{\hfil/\hfil}}      % so center / in box
      #1                                        % and print #1
   \else                                        % / is bigger
      \rlap{\hbox to \dimen1{\hfil$#1$\hfil}}   % so center #1
      /                                         % and print /
   \fi}                                         %

\begin{document}
\title{Triangularity and Dipole Asymmetry in Heavy Ion Collisions}
\author{Derek Teaney and Li Yan}
\affiliation{Department of Physics and Astronomy, Stony Brook University, Stony Brook, New York 11794-3800, United States}
%\author{Kevin Dusling}
\date{\today}

\begin{abstract}
We introduce a cumulant expansion to parameterize possible initial conditions
in relativistic heavy ion collisions. We show that the cumulant expansion
converges and that it can systematically reproduce the results  of Glauber type
initial conditions.  At third order in the gradient expansion, the cumulants
characterize the triangularity $\llangle r^3 \cos3(\phi - \psi_{3,3} )
\rrangle$ and the dipole asymmetry $\llangle  r^3 \cos(\phi- \psi_{1,3})
\rrangle$ of the initial entropy distribution.  We show that for mid-peripheral
collisions the orientation angle of the dipole asymmetry $\psi_{1,3}$ has a
$20\%$ preference out of plane.  This leads to a  small net $v_1$ out of plane.
In peripheral and mid-central collisions  the orientation angles $\psi_{1,3}$
and $\psi_{3,3}$ are strongly correlated, but this correlation disappears
towards central collisions.   We study the ideal hydrodynamic response to these
cumulants and determine the associated $v_1/\epsilon_1$ and $v_3/\epsilon_3$
for a massless ideal gas equation of state.  The space time development of
$v_1$ and $v_3$  is clarified with figures. These figures show  that $v_1$ and
$v_3$ develop towards the edge of the nucleus,   and consequently the final
spectra are more sensitive to the viscous dynamics of freezeout.  The
hydrodynamic calculations for $v_3$ are provisionally compared  to Alver and
Roland fit of STAR inclusive two particle correlation functions.  Finally, we
propose to measure the $v_1$ associated with the dipole asymmetry  and the
correlations between $\psi_{1,3}$ and $\psi_{3,3}$ by measuring  a two particle
correlation with respect to the participant plane, $\llangle \cos(\phi_\alpha -
3\phi_\beta  + 2\Psi_{PP}) \rrangle $.   The hydrodynamic prediction for this
correlation function is several times larger than a correlation currently
measured by the STAR collaboration, $\llangle \cos(\phi_\alpha + \phi_\beta -
2\Psi_{PP} ) \rrangle$.  This experimental measurement would provide convincing
evidence for the hydrodynamic and geometric interpretation of two particle
correlations at RHIC.
\end{abstract}

\maketitle
\section{Introduction}

In a recent  and significant paper B.~Alver and G.~Roland (AR)
\cite{Alver:2010gr} provided the most compelling explanation to date for the
striking features measured  in two particle correlations  at the Relativistic
Heavy Ion Collider \cite{Adams:2005ph,Adare:2007vu,Alver:2009id,Abelev:2009qa,Adamova:2009ah,Abelev:2008nda}.  These features (which are described with
picturesque  names such as the ``ridge" and ``shoulder")  are said to arise from 
the collective response to fluctuating initial conditions.  Specifically,
if the initial conditions are parameterized with 
a quadrapole and triangular moment, the 
two particle correlations reflect the response of the nuclear medium to 
these anisotropies. The work of AR was motivated in part by event by event simulations
of heavy ion collisions with ideal hydrodynamics
which  showed that the flow from fluctuating 
initial conditions
 can describe the general features of the measured two particle
correlations \cite{Takahashi:2009na}.  The general idea that the curious 
correlations  are due to a third harmonic in the flow profile was previously
suggested by Sorensen  \cite{Sorensen:2010zq}. In addition, many of the features of the observed two particle correlations were found in the AMPT model 
\cite{Lin:2004en,Ma:2006fm, Zhang:2007qx}, 
though the geometric nature of these correlations was not understood before
the work of Alver and Roland.

The hydrodynamic interpretation of the measured two particle correlations is
important for several reasons.  First, before this conclusion there was
a significant  correlation  between the measured particles which was not
understood. This confusion casted doubt on the  hydrodynamic interpretation of
RHIC results and clouded the important conclusion that the shear viscosity to
entropy ratio of QCD is of order $\sim \hbar/4\pi$ near the phase transition
\cite{Teaney:2009qa}.  However, since the unusual two particle correlations are
actually a prediction of hydrodynamics,  the observation of these  unusual
features in the data validates hydrodynamics as an appropriate effective theory
for heavy ion events and marginalizes other models.  Further, once the
hydrodynamic interpretation is adopted the measured correlations can be used to
constrain the properties of the medium, {\it e.g.} the shear viscosity and the
Equation of State (EOS).  In particular the effect of viscosity was calculated in  Refs.~\cite{Alver:2010dn,Schenke:2010rr}  which will be discussed more completely below.  

Motivated by these results, the current work will characterize the
fluctuating initial conditions with a cumulant expansion.  Instead of running
hydrodynamics event to event, the linear response to specified  cumulants can
be calculated with ideal or viscous hydrodynamics.  Subsequently, these
response functions can be combined with a Glauber model  for the event-by-event
cumulants (and their correlations), and the combined result can be fairly
compared to data.  At third order in the gradient expansion,  the initial
condition is parameterized by  a radial dependence to the  dipole moment,
$\llangle r^3 \cos\phi \rrangle$, and the triangularity, $\llangle r^3
\cos3\phi \rrangle$\,.  In \Sect{time} we will calculate (with ideal
hydrodynamics) how the medium responds to these moments  and illustrate how
this response develops  in space and time. Subsequently in \Sect{spectra} we
will compute the corresponding particle spectra $v_1(p_T)$ and $v_3(p_T)$ and
study the sensitivity to certain model parameters related to freezeout.
In \Sect{conclusion_a} we will make a comparison  to $V_{3\Delta}/V_{2\Delta}$ as
extracted by Alver and Roland in their analysis of two particle correlations. 
We will also make definite predictions for the dipole asymmetry $v_1(p_T)$,
which, if confirmed, would firmly establish the geometric nature of the two
particle correlations.
The comparison to data is not final as the effects of resonance 
decays, viscosity, and higher  cumulants  have not been included, 
Nevertheless, the preliminary comparison  will firmly tie the formal
cumulant expansion  outlined in this paper to the measured correlations.
Finally, we will compare our calculations to the recent results of 
Refs.~\cite{Petersen:2010cw,Alver:2010dn, Schenke:2010rr} in \Sect{conclusion_b}.

\section{Cumulant expansion and hydrodynamics at RHIC }
\label{cumulant}

\subsection{The initial conditions for ideal hydrodynamics}

In this paper we will use $2+1$ dimensional boost invariant ideal hydrodynamics
to simulate RHIC events \cite{Ollitrault:1992bk, Kolb:2003dz, Teaney:2009qa}. 
Briefly, in ideal hydrodynamics the stress tensor 
satisfies the constituent relation 
and the conservation laws:
\st
 T^{\mu\nu}  = (e  + \pr(e)) u^{\mu} u^{\nu} + \pr(e) g^{\mu\nu}\, ,   
 \qquad \nabla_{\mu} T^{\mu\nu} = 0  \, , 
\stp
where $e$ is the energy density, $u^{\mu}$ is 
the flow velocity, and the  pressure $\pr$ is specified 
by the EOS, $\pr = \pr(e)$.
We will work in flat space 
but with coordinates
\[
\tau = \sqrt{t^2 - z^2} \, , 
\qquad \eta_s = \frac{1}{2} \log\left( \frac{t+z}{t-z} \right) \, .
\]  
With the assumption of boost invariance,  the hydrodynamic fields are
independent of $\eta_s$ and $u^{\eta}=0$.  Using the constraint 
$u_{\mu} u^{\mu} = -1$,  the independent fields which  must
be determined by solving the  conservation laws are 
\st
   e(\tau, \x)\, ,   \qquad  u^{x}(\tau, \x) \,  \qquad u^{y}(\tau, \x)  \, ,
\stp
where $\x$ denotes two dimensional vectors in the transverse plane. 
We will specify the initial conditions for the subsequent evolution 
in what follows. At the initial time $\tau_o$ it is reasonable to assume
that flow fields are small,  $u^{x} \simeq u^{y} \simeq 0$. 
This leaves the initial energy density which must be specified $e(\tau_o, x, y)$. 
We will specify the initial entropy density $s(\tau_o, x, y)$ with a cumulant 
expansion and infer the initial energy density from the equation of state. 

%\Fig{typical} illustrates a typical initial 
%condition and the general ideal of the cumulant expansion.
A typical initial condition might be fairly complicated involving
several structures. However, the effect of the shear viscosity 
is to damp the highest Fourier modes.  Thus, after damping the shortest
wavelengths, the initial entropy distribution is approximately
described by a Gaussian with average squared radius $\llangle r^2 \rrangle$
and elliptic eccentricity  $\epsilon_2$ as has traditionally been used to 
characterize
heavy ion events \cite{Ollitrault:1992bk}.  The damping of the highest Fourier modes  is nicely seen 
in Fig.~1  of a recent preprint \cite{Schenke:2010rr}.  The next paragraphs formalize 
this description and categorize corrections.

\subsection{Cumulants}

The Fourier transform of the entropy density  for a given initial
condition is 
\st
\label{fourier}
  \int \dd^2\x \, e^{i\k\cdot\x } \rho(x)  = \rho(\k) \,  ,
\stp
where $\rho(\x) = \tau_o s(\tau_o,\x)/S_{\rm tot}$  and $S_{\rm tot} = 
\int \tau_o \dd^2\x \, s(\tau_o,\x)$ is the total
entropy per space time rapidity.   Since the highest Fourier 
modes are damped, we will expand the initial distribution in $\k$.
Expanding both sides  of \Eq{fourier} with respect to $\k$,
\st
\rho(\k)  = 1 +  i k^{a} \rho_{1,a} +  \frac{(i k^{a}) (i k^b)}{2!} \rho_{2,ab} + \ldots \, ,
\stp
we see that  $\rho(\k)$ generates moments of the entropy distribution
\st
 \rho_{1,a} = \llangle x_{a} \rrangle\, , \qquad \rho_{2,ab} = 
\llangle x_{a} x_{b} \rrangle \, ,
\stp
where the average is appropriately defined
\st
\label{avedef}
 \llangle \ldots \rrangle =  \int \dd^2\x \rho(\x) \ldots \, .
\stp
Although we could  classify the initial conditions with these  moments,
a cumulant expansion seems more natural since the average
Glauber distribution is roughly  Gaussian and the cumulants are 
translationally invariant.
We therefore define $W(\k)$  
\st
 \exp(W(\k)) \equiv \int \dd^2\x \, e^{i\k\cdot \x} \rho(\x)  \, , 
\stp
and  expand both sides in a fourier series
\st
 W(\k)  = 1 + i k^{a} W_{1,a} + \frac{1}{2!} (i k^a)(ik^a)  W_{2,ab} + \ldots \, .
\stp
From this expansion we see that $W(\k)$ is the generating function 
of cumulants of the underlying distribution $\rho(\k)$
\st
 W_{1,l} = \llangle x_l \rrangle \, ,   \qquad W_{2,ab} = \llangle x_a x_b \rrangle - \llangle x_a \rrangle  \llangle  x_b \rrangle \, .
\stp 
From now on we will  shift the origin so that $\llangle x_a \rrangle = 0$, 
and the distribution is approximately  Gaussian to quadratic order
\st
  \rho(\k) =  \exp\left( - \frac{1}{2} k^a k^b W_{2,ab} \right) \, . 
\stp
Higher order corrections in this expansion will correct the distribution
away from the Gaussian. 
The tensor $W_{2,ab}$ is a reducible tensor and 
should be decomposed into irreducible components,
\st
  W_{2,ab} = \frac{1}{2} W_{2,cc} \delta_{ab}   + \left(W_{2,ab} - \frac{1}{2} W_{2,cc} \delta_{ab} \right) \, .
\stp   
We  orient the  $x,y$  axes to the participant plane \cite{Alver:2006wh} 
where $W_{2,xy} = 0$. Then the irreducible moments are 
\begin{align}
W_{2,aa} =& \llangle x^2 + y^2 \rrangle \, , \\
W_{2,xx} - \frac{1}{2} W_{cc} \delta_{xx} 
 &= \frac{1}{2} \llangle x^2 - y^2 \rrangle \, .   
\end{align}
Clearly the irreducible components of the cumulant expansion are  related
to the traditional parameters of heavy ion physics:
\begin{align}
\llangle x^2 + y^2 \rrangle \, ,  \qquad \mbox{and} \qquad 
\epsilon_{2} \equiv \frac{\llangle  y^2 - x^2 \rrangle}{r^2} \, .
\end{align}

To write down  corrections to these results it is more convenient and 
illustrative
to use cylindrical tensors  rather than  Cartesian tensors.
\App{details} develops this expansion in detail and 
only certain features will be summarized here. \App{cumulants_formal} 
expands $W(\k)$ in 
a Fourier series:
\st
 W(\k) = W_0(k) + 2 \sum_{n=1}^{\infty} W_{n}^c(k) \cos(\phi_k) + 2 \sum_{n=1}^{\infty}  W_{n}^s (k) \sin(n \phi_k) \, ,
\stp
where $k$ and $\phi_\k$ are the norm and azimuthal angle of 
the momentum vector. 
The $W_{n}^{c,s}(k)$  are also expanded in $k$ to characterize the distribution 
at largest wavelength:  
\begin{subequations}
\label{Wnk_1}
\begin{align}
  W_{0}(k) =&  \frac{1}{2!} W_{0,2}(ik)^2  + O(k^4) \, ,   \\
  W_{1}^c(k) =& W_{1,1}^c  + O(k^3) \, , \\
  W_{1}^s(k) =& W_{1,1}^s  + O(k^3) \, , \\
  W_{2}^c(k) =& \frac{1}{2!} W_{2,2}^c (ik)^2  +  O(k^4) \, , \\
  W_{2}^s(k) =& W_{2,2}^s  + O(k^4) \, .
\end{align}
\end{subequations}
After \App{details} we find   that to order $k^2$
\begin{align}
  W_{0,2}   =& \frac{1}{2} \llangle r^2 \rrangle \,  , \\
  W_{1,1}^c =&  0 \, , \\
  W_{1,1}^s =&  0 \, , \\
  W_{2,2}^c =& \frac{1}{4} \llangle r^2 \cos(2\phi) \rrangle \, , \\
  W_{2,2}^s =&  0  \, .
\end{align}
Here we have used translational invariance and rotational invariance (as in the
Cartesian case) to eliminate $W_{1,1}^c$, $W_{1,1}^s$, and $W_{2,2}^s$.
To third order in the gradient expansion  the dipole terms $W_{1}^c(k)$ 
and $W_{1}^s(k)$ are non-zero
\begin{subequations}
\label{Wnk_2}
\begin{align}
  W_{1}^{c}(k) =&  \frac{1}{3!} W_{1,3} (ik)^3  + O(k^5) \, ,  & W_{1,3} =& \frac{3}{8} \llangle r^3\cos\phi \rrangle  \, ,  \\
  W_{1}^{s}(k) =&  \frac{1}{3!} W_{1,3} (ik)^3  + O(k^5) \, , & W_{1,3} =& \frac{3}{8} \llangle r^3\sin\phi \rrangle   \, .
\end{align}
Similarly, at third order in the gradient expansion there are terms
proportional to $\cos(3\phi)$
\begin{align}
  W_{3}^{c}(k) =&  \frac{1}{3!} W_{3,3} (ik)^3 + O(k^5) \, ,   &   W_{3,3}^c =& \frac{1}{8} \llangle r^3 \cos(3\phi) \rrangle \, , \\
  W_{3}^{s}(k) =&  \frac{1}{3!} W_{3,3} (ik)^3 + O(k^5)    &  W_{3,3}^s =&  \frac{1}{8} \llangle r^3 \sin(3\phi) \rrangle  \, .
\end{align}
\end{subequations}
Once the fourier coefficients {$W_{n,m}$} are specified, the entropy 
distribution in space can be  found with a fourier transform; 
see Eqs.~\ref{triangle_space} and \ref{dipole_space} and the surrounding text for further discussion.

\subsection{A strategy for event by event hydrodynamics} 
If the cumulants beyond second order are in some sense small,  
then the change in the 
hydrodynamic spectra due to a specified set of higher cumulants is linearly 
proportional to the deformation
\st
  \frac{\dd \delta N}{\dd\phi_p } 
=   \sum_{n ,m, \left\{s,c\right\} } \left[ \frac{1}{W_{n,m}^{c,s}}   \frac{\dd  \delta N}{\dd \phi_p}  \right]_{n,m,\left\{s,c\right\}}  W_{n,m}^{c,s}  \; ,
\stp 
where 
\st
\left[ \frac{1}{W_{n,m}^{c,s}}   \frac{\dd \delta N}{\dd\phi_p}  \right]_{n,m,\left\{c,s\right\}}  \, ,  
\stp
is the normalized response to a given cumulant.  If the non-linear interactions
between the elliptic flow and the higher cumulants can be ignored ({\it i.e.}  the 
elliptic flow is sufficiently small), then the background Gaussian is approximately 
radially symmetric and the response of the $\sin$ terms are related to the response 
of the cosine terms through a rotation.  In this case, we are free to rotate our coordinate  
system by an angle $\psi_{n,m}$ 
\st
  \psi_{n,m} = \frac{1}{n} {\rm atan2}( W_{n,m}^s, W_{n,m}^c ) + \frac{\pi}{n} \, ,
\stp
so that the $\sin$ terms vanish. 
In this rotated frame (which we will notate as $\hat{W}$) the cumulants are 
\st
  \hat{W}_{n,m}^s = 0\,  ,  \qquad \hat{W}_{n,m}^c = - \sqrt{(W_{n,m}^c)^2 + (W_{n,m}^s)^2 } \, ,
\stp
and  the spectrum can  be written
\st
 \frac{\dd \delta N}{\dd\phi_p } 
=   \sum_{n ,m, c } \left[ \frac{1}{\hat W_{n,m}^{c}}   \frac{\dd \delta N}{ \dd (\phi_p - \psi_{n,m} )} \right]_{n,m,\left\{c\right\}}  \hat{W}_{n,m}^{c}  \, .
\stp
Thus, the assumption of a rotationally 
invariant background reduces the number of coefficients by a factor of two.

In this paper we will assume that all deformations from spherical are small
including the elliptic flow. Thus, we will neglect the non-linear couplings
between the elliptic flow and the triangular flow and the elliptic flow and the
dipolar flow.  We have investigated the influence of the ellipticity on the
triangular and dipolar flow and our preliminary findings show that  the effect
of the elliptic flow on the triangular flow is small.  A similar finding 
was reported  in the very recent preprint  by the Duke group \cite{Qin:2010pf}. 
However, the effect of
the elliptic flow on the dipolar flow is non-negligible when the dipole angle
is oriented in plane.  This complication will be reported on in future work
\cite{YanWork}. 

The angle $\psi_{2,2}$ specifies the orientation 
of the participant plane $\Psi_{PP}$, and 
the second order 
cumulant $\hat W_{2,2} $  determines
the  ellipticity 
\begin{align}
 \epsilon_{2} \equiv - \frac{ \llangle r^2 \cos2(\phi - \Psi_{PP}) \rrangle  }{\llangle r^2 \rrangle } = -\frac{4\hat W_{2,2}^c }{\llangle r^2 \rrangle} \, , 
 \qquad  \Psi_{PP} \equiv \psi_{2,2} \, .
\end{align}
 The participant plane angle $\Psi_{PP}$  is distinct from the 
reaction plane angle which we  denote with $\Psi_R$. 

The third order cumulant $\hat W_{3,3}^c$ 
describes the triangularity as introduced by Alver and Roland.
These authors suggested a definition of
the triangularity and orientation angle $\epsilon_3^{AR}$ and $\psi_3^{AR}$
with a quadratic radial weight
\st
 \epsilon_3^{AR} = -\frac{\llangle r^2 \cos(3(\phi - \Psi_3^{AR})) \rrangle }{\llangle r^2 \rrangle } \, ,
 \qquad \Psi_3^{AR} = \frac{1}{3} {\rm atan}2( \llangle r^2 \sin(3\phi) \rrangle, \llangle r^2 \cos(3\phi) \rrangle )  + \frac{\pi}{3} \, . 
\stp 
We will abandon this analytically frustrated definition, and 
define the triangularity $\epsilon_3$ and the associated angle 
with an $r^3$ weight 
\begin{align}
\label{epsilon_3def}
 \epsilon_{3}  \equiv& 
   - \frac{ \llangle r^3 \cos3(\phi - \psi_{3,3}) \rrangle }{\llangle r^3\rrangle}  = -\frac{8\hat{W}_{3,3}^c }{ \llangle r^3 \rrangle} \, ,
\\
   \psi_{3,3} \equiv& \frac{1}{3} {\rm atan}2(\llangle r^3\sin3\phi \rrangle, \llangle r^3 \cos3\phi \rrangle )  + \frac{\pi}{3} \, .
\end{align}
The difference between the $r^2$ and $r^3$ weight is captured by the response
of the system to the fifth order cumulants, $W_{3,5}^{c} \propto \left[\llangle r^5 \cos 3\phi \rrangle - 4\llangle r^2 \rrangle \llangle r^3 \cos 3\phi \rrangle \right] $ . 
Recent studies of the response of the system to $\epsilon_{5}$ (or
$W_{5,5}$ in the current context) suggests that 
the response to these fifth order cumulants will be 
small \cite{Alver:2010dn}. 

The third order cumulant $\hat W_{1,3}^c$ 
describes a dipole asymmetry and 
also appears to the same order in the gradient expansion. By analogy we define
$\epsilon_1$ and $\psi_{1,3}$
\begin{align}
   \epsilon_{1} \equiv&  -\frac{\llangle r^3 \cos(\phi - \psi_{1,3})  \rrangle }{\llangle r^3 \rrangle} = -\frac{8}{3} \frac{\hat{W}_{1,3}^c}{\llangle r^3 \rrangle }  \, , \\
   \psi_{1,3} \equiv&  {\rm atan}2 ( \llangle r^3\sin \phi \rrangle, \llangle r^3\cos\phi \rrangle) + \pi  \, .
\end{align}
Estimates for these parameters and their correlations will  
be given in the next section.

\subsection{ The dipole asymmetry and triangularity }
\label{dipole_tripole}

\begin{figure}
\begin{center}
\includegraphics[width=0.33\textwidth]{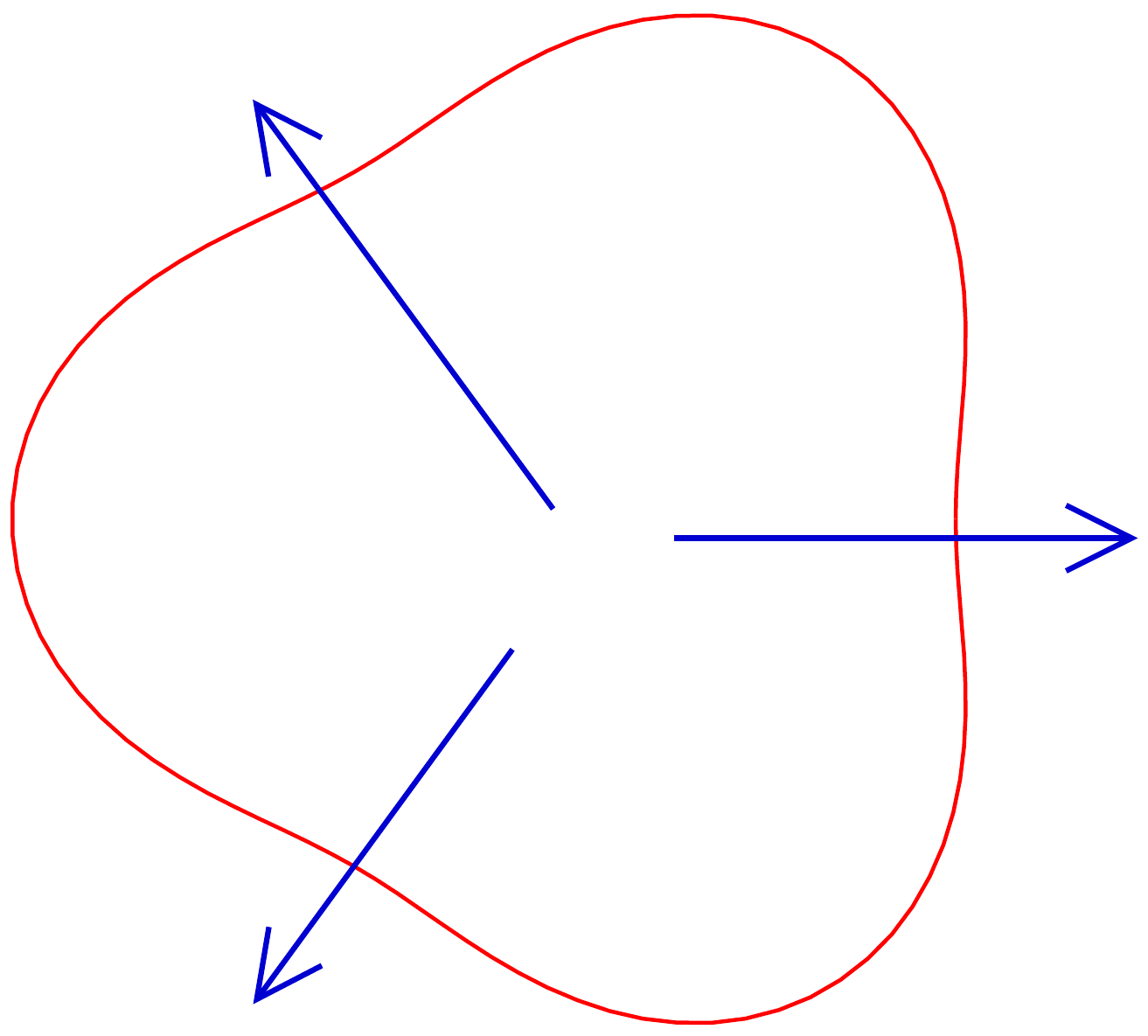}
\hspace{0.4in}
\includegraphics[width=0.33\textwidth]{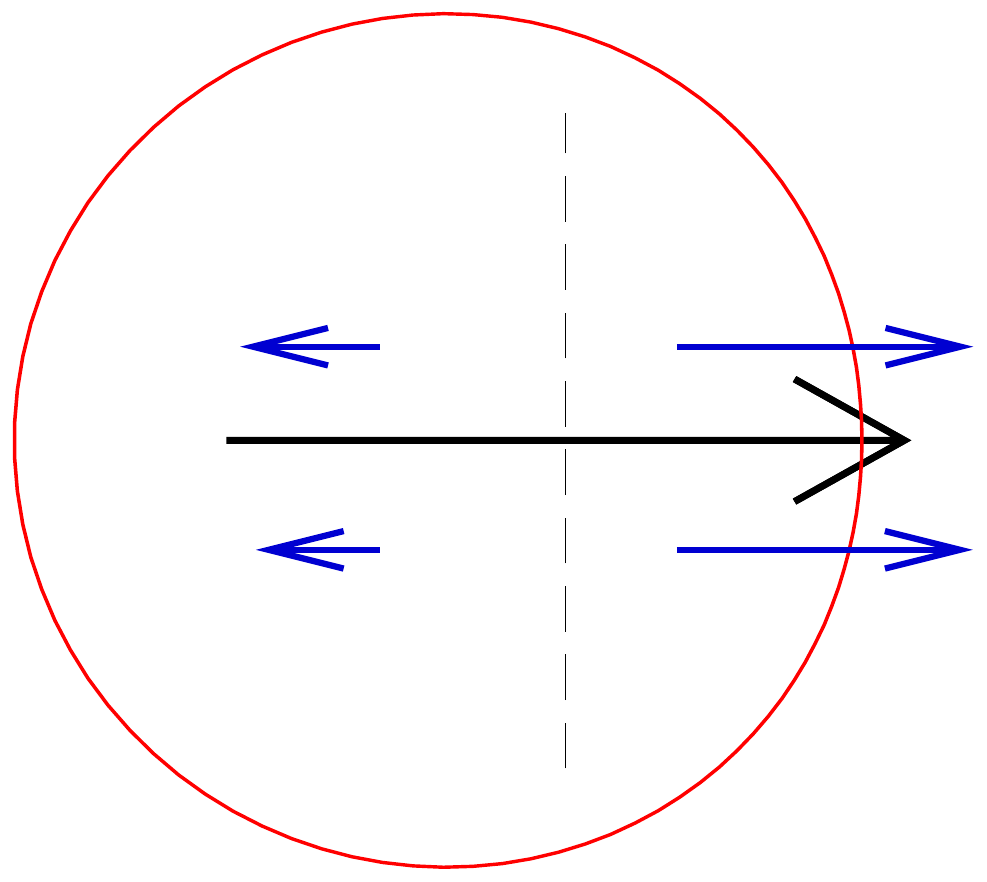} 
\end{center}
\caption{A schematic of an event with (a) net triangularity 
and (b) net dipole asymmetry. The triangularity produces a net $v_3(p_T)$ 
and the dipole asymmetry produces a net $v_1(p_T)$. The cross in
(b) indicates the center of entropy (analogous to the center of mass)
and the large arrow indicates the orientation of the dipole.
\label{sample_initfig}}
\end{figure}

To get a feeling for the dipole asymmetry and 
triangularity we first record the explicit coordinate space
expressions for a distribution with only triangularity
\st
\label{triangle_space}
s(\x,\tau)   \propto   \left[1 +  \frac{\llangle r^3 \rrangle \epsilon_3}{24}  \left( \left(\frac{\partial }{\partial x} \right)^3 - 3 \left(\frac{\partial}{\partial y}  \right)^2 \frac{\partial}{\partial x} \right) \right] e^{- \frac{r^2}{\llangle r^2 \rrangle } }
 \, , 
\stp
and a distribution with only a dipole asymmetry
\st
\label{dipole_space}
s(\x,\tau)   \propto  
 \left[ 1 + \frac{ \llangle r^3 \rrangle \epsilon_1 }{8 } \left( \left(\frac{\partial}{\partial x} \right)^3  + \left( \frac{\partial}{\partial y} \right)^2 \frac{\partial}{\partial x}  \right) \right] e^{- \frac{r^2}{\llangle r^2 \rrangle }}
\, . 
\stp
Here the orientation angles $\psi_{3,3}$ and $\psi_{1,3}$ are set to zero.
At large enough radius the derivative terms become large 
and overwhelm the leading term making the distribution negative.
This is an unavoidable consequence of truncating a cumulant expansion 
at any finite order.  
As explained in \App{details} we regulate these terms 
and adjust the overall constant to reproduce the total 
entropy in a central RHIC collision.
\Fig{sample_initfig}a and \Fig{sample_initfig}b illustrate
initial conditions with net triangularity and net dipole asymmetry respectively.
The distribution with net triangularity
leads to a $v_{3}(p_T)$ while the dipole asymmetry leads to a $v_1(p_T)$.

\begin{figure}
\includegraphics[height=3.0in]{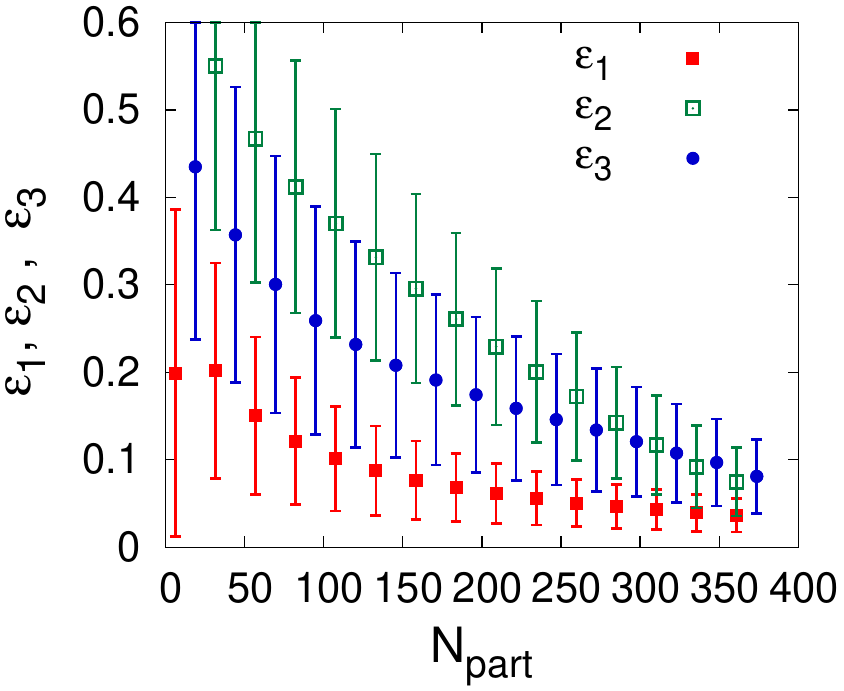}
\caption{ Size of the moments  $\epsilon_1$, $\epsilon_2$ and $\epsilon_3$ as a function of the number of participants. The points indicate 
the average value of $\dlangle \epsilon_n \drangle$ and the errorbars indicate the 
variance of $\epsilon_n$  at fixed $N_{\rm part} $.  }
\label{envsnpart} 
\end{figure}

To estimate these parameters and their correlations we have
used the PHOBOS monte carlo Glauber code \cite{Alver:2008aq}.
\Fig{envsnpart} shows the distribution 
of $\epsilon_1$, $\epsilon_2$ and $\epsilon_3$ as a function of the number
of participants. 
We see that the dipole asymmetry is  about a factor of 
two smaller than the triangularity but is not negligibly small.

\begin{figure}
\includegraphics[width=6.5in]{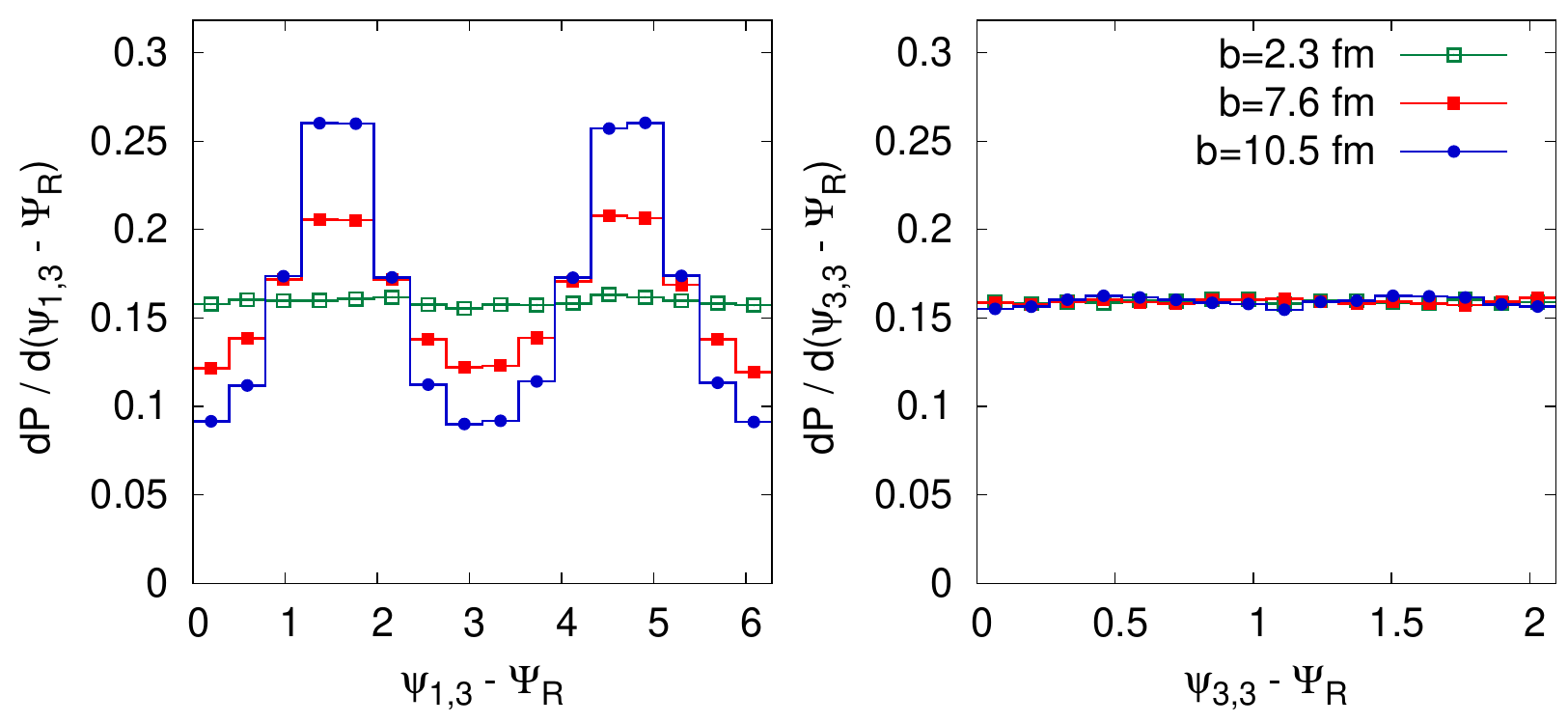}
\caption{
Distribution of the angles $\psi_{1,3}$ and $\psi_{3,3}$ with respect to the reaction plane for three different impact parameters.  \label{angledist}
}

\end{figure}

\Fig{angledist} shows the distribution of $\psi_{1,3}$ and $\psi_{3,3}$ 
with respect to reaction plane at various 
impact parameters. We see that although $\psi_{3,3}$
is uncorrelated with respect to the reaction plane, $\psi_{1,3}$ 
shows an anti-correlation with respect to the reaction plane,
which eventually disappears toward central collisions.  
% The
% amplitude of the variation in the mid-peripheral bin is of 
% order 

\begin{figure}
\begin{center}
\includegraphics[width=3.5in]{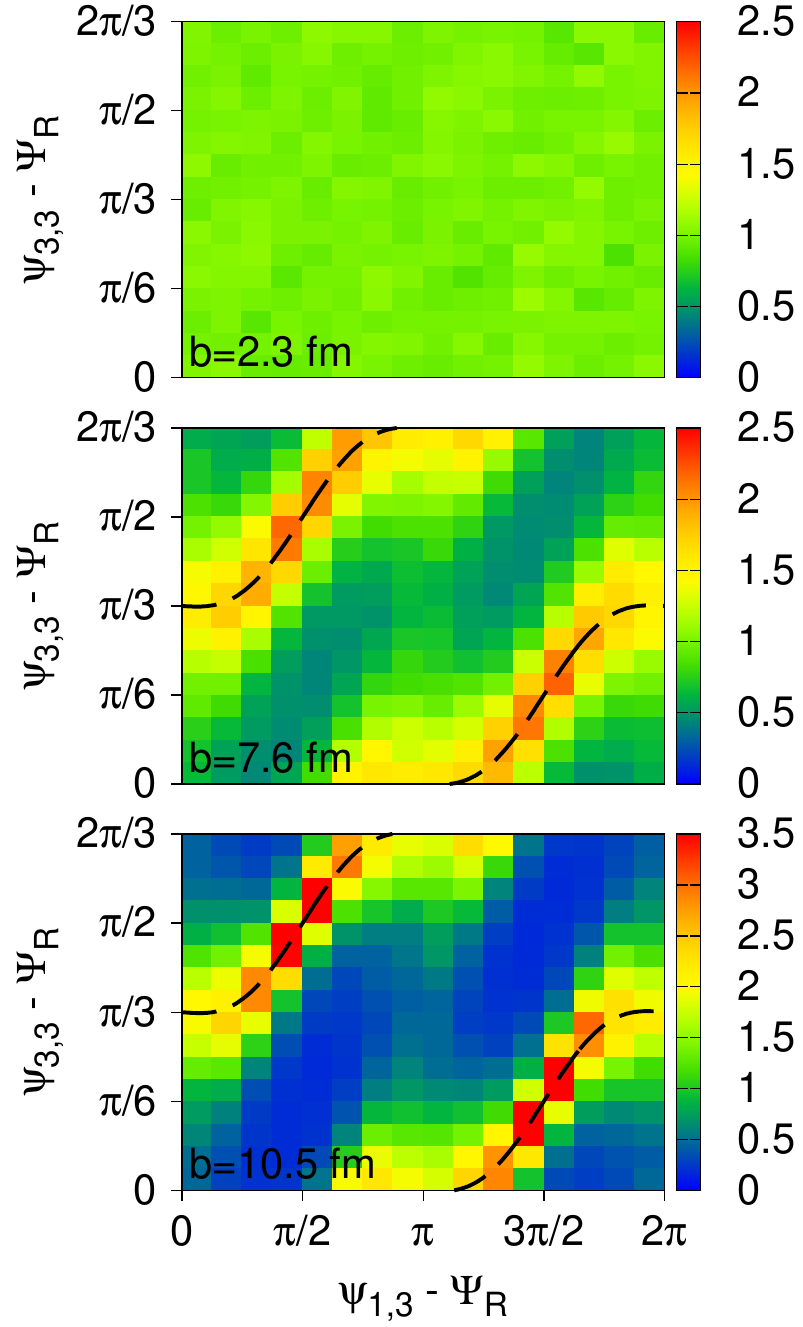}
\end{center}
\caption{
The conditional probability distribution $P(\psi_{3,3} | \psi_{1,3} \Psi_R)$  
for three different impact parameters $b=0,7.6, 10.5$ fm. The functional 
form of the dashed curve is given by \Eq{phase} with 
fit parameter $C = 0.53$ for $b=7.6\,{\rm fm}$ and $C=0.56$ for $b=10.5\,{\rm fm}$. 
\label{correlation_fig}
}
\end{figure}

\begin{figure}
\includegraphics[width=\textwidth]{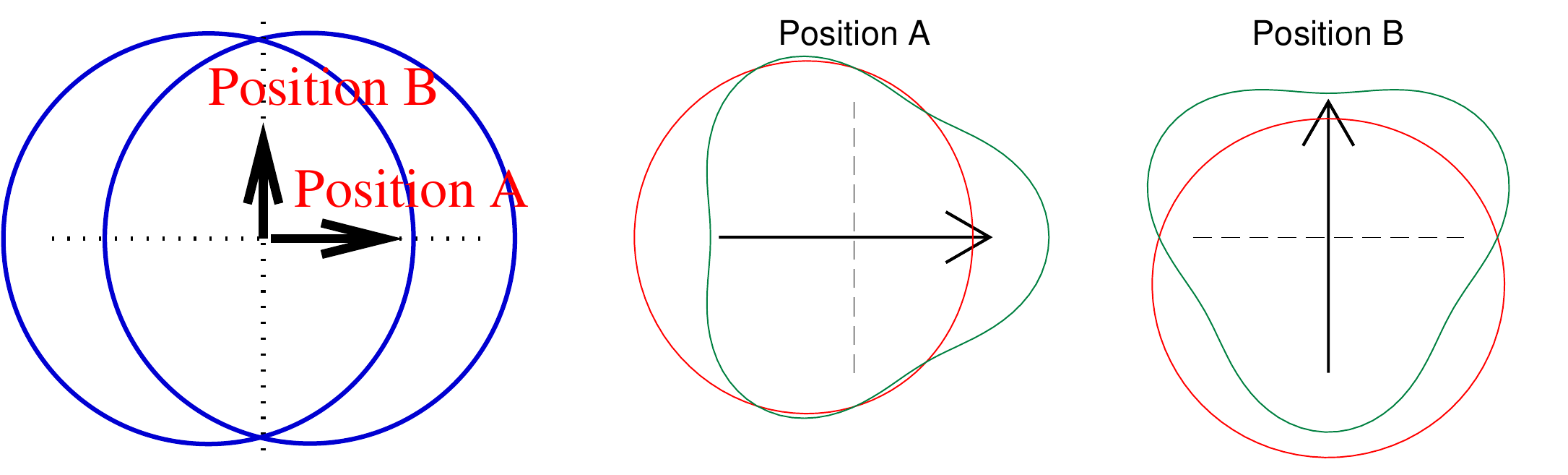}
\caption{
The figure qualitatively describes the fluctuations  associated with the Glauber 
model as illustrated in Fig.~\protect\ref{correlation_fig}. When the dipole asymmetry is in plane (Position A), then the 
tip of triangularity is aligned with dipole asymmetry. When
the dipole asymmetry is out of plane (Position B), the tip of the triangle
is anti-aligned with the dipole asymmetry.
\label{corrfig}
}
\end{figure}

More importantly, the angles $\psi_{1,3}$ and $\psi_{3,3}$ are
strongly correlated  in mid central collisions (a similar observation was made recently by Staig and Shuryak \cite{Staig:2010pn}).
\Fig{correlation_fig} shows the conditional probability distribution, {\it i.e.}
\[
 P(\psi_{3,3} | \psi_{1,3},\Psi_R )  \equiv \mbox{The probability of $\psi_{3,3}$ given $\psi_{1,3} $  and $\Psi_R$}.
\]
The strong correlation may be explained physically as follows. When
the dipole asymmetry is in plane then the triangular axis is at $\pi/3$,
{\it i.e.} the point of the triangle is aligned with the dipole axis as
exhibited in \Fig{corrfig}(a).
However, when the dipole axis is out of plane then the triangular axis
is also out of plane as exhibited in \Fig{corrfig}(b). 

These correlations
are a reflection of the almond shape geometry and their general form can be 
established  by symmetry arguments. 
First, since the probability of finding 
a dipole asymmetry in a given quadrant of the ellipse is the same for 
every quadrant, the probability distributions $\dd P/\dd (\psi_{1,3} - \Psi_{R})$ 
must only involve even cosine terms
\st
\label{Acoeffdef}
 \frac{\dd P}{\dd \psi_{1,3} }  = \frac{1}{2\pi} \left(1 -  2 A \cos 2(\psi_{1,3} - \Psi_R)) + \ldots \right) \, .
\stp
The sign has been chosen so that a positive $A$ coefficient describes
the out-of plane preference seen in \Fig{angledist}. 
The coefficient $A$ must vanish in a  cylindrically symmetric collision,
and for small anisotropy we have 
\st 
A \propto \aveeps \, ,
\stp 
where  
the double brackets denotes an event averaged $\epsilon_2$
Similarly $\dd P/\dd(\psi_{3,3} - \Psi_R)$ must involve even cosine 
terms and must be $2\pi/3$ periodic
\st
 \frac{\dd P}{ \dd(\psi_{3,3} - \Psi_R) } = \frac{1}{2\pi} 
\left(1 + 2 A_6\cos(6(\psi_{3,3} - \Psi_R))  + \ldots \right) \, .
\stp
The relatively high fourier number $n=6$ explains the smallness of 
the observed asymmetry, and $A_6$ will be ignored from now on. 
The form of the conditional probability distribution can also be 
established based on general considerations. \App{glauber} uses symmetry 
arguments, a fourier expansion, and the statement that the correlation is 
strongest when the triangle and dipole angles are aligned at $\pi/2$ out of 
plane, to establish a three parameter functional form  which describes
the correlations fairly well 
\st
\label{p1p3pr}
 P(\psi_{3,3}|\psi_{1,3}, \Psi_R)
 = 
\frac{1}{2\pi} \Big[ 1 - 
 2  \left( B_0   -  2 B_2 \cos(2\psi_{1,3} - 2\Psi_R) \,  \right)  \cos(3 \psi_{3,3} - \phi^{*} - 2\Psi_R)    \Big] \, ,
\stp
where 
\st
\label{phase}
\phi^*=\psi_{1,3} - C\sin(2\psi_{1,3} - 2\Psi_{R})  \, .
\stp
The signs are chosen so that
$B_0$, $B_2$, and $C$ are positive constants in the final fits.
 A sample fit with this functional form is given in \Fig{fitplot} 
of \App{glauber}.
The phase angle $\phi^*$ is illustrated by the  dashed black line in \Fig{correlation_fig},
 which is found by solving   $3\psi_{3,3} - \phi^{*} + 2\Psi_R = \pi$ for $\psi_{3,3}$.
Although we have written the conditional probability when the reaction plane angle is fixed, 
the same arguments could have been used to determine the 
functional form of the conditional probability when  the participant
plane angle is fixed, {\it i.e.  }
\st
 P(\psi_{3,3}|\psi_{1,3} \Psi_{PP})
 =  \mbox{\Eq{p1p3pr} with $\Psi_{R}  \rightarrow \Psi_{PP}$  and sightly different numerical coefficients}. 
\stp

In the limit of small elliptic eccentricity the coefficients scale as 
\st
\label{epsscaling}
 B_0 \propto \aveeps \, ,  \qquad B_2 \propto \aveeps^2 \, ,  \qquad C \propto \aveeps \, ,
\stp
as is shown in \App{glauber}.
%We have inserted powers of $\bar \epsilon \equiv \llangle \epsilon_2 \rrangle$ 
%in front of the coefficients so that the coefficients are finite in the limit 
%$\epsilon \rightarrow 0$. 
Thus  the conditional probability distribution  simplifies 
in this limit to
\st
\label{B0coeffdef}
P(\psi_{3,3} | \psi_{1,3} \Psi_R)
 = \frac{1}{2\pi}
 \left[ 1 -  2 B_0 \cos(3 \psi_{3,3} -\psi_{1,3} - 2\Psi_{R} ) \right],
\stp 
which describes almost all of the essential physical features. 

\begin{figure}
\includegraphics[height = 2.5in]{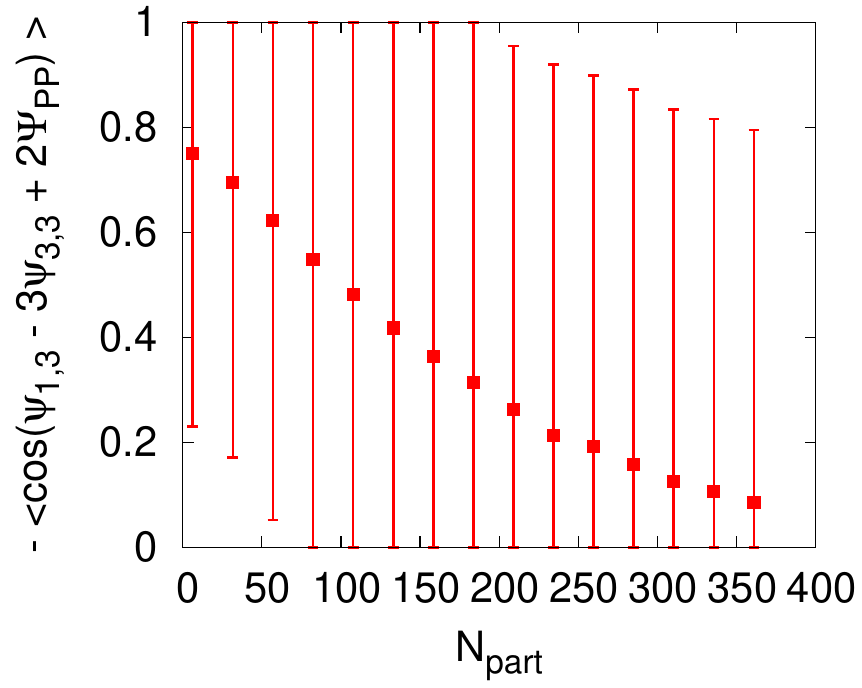} 
\caption{
Correlation of the true dipole angle $\psi_{1,3}$  and the estimated event plane 
angle $\psi_{1,3}^{\rm mp} = 3\psi_{3,3} - 2 \Psi_{PP} - \pi$. The points indicate
the average $\dlangle \cos(\psi_{1,3} - \psi_{1,3}^{\rm mp}) \drangle$  
and the errorbars indicate the variance  of this quantity at fixed $N_{\rm part}$. 
\label{cospest}
}
\end{figure}

The strong correlation means that if the triangular and the participant planes  are known,
then the dipole plane can be determined statistically.
The probability distribution of $\psi_{1,3}$ for fixed $\psi_{3,3}$ and $\Psi_{PP}$ is approximately
\begin{multline}
 P(\psi_{1,3} | \psi_{3,3}\Psi_{PP})
 \simeq \frac{1}{2\pi}  \Big[ 1 + 2 A \cos(2\psi_{1,3} - 2\Psi_{PP} )  
-  2 B_0  \cos(3\psi_{3,3} - \psi_{1,3} - 2\Psi_{PP} ) \Big] \, .
\end{multline}
Maximizing this probability we determine the most probable angle of $\psi_{1,3}^{\rm mp}$ given $\psi_{3,3}$ and $\Psi_{PP}$.
Neglecting the  $A$ coefficient  which is significantly 
smaller than $B_0$ we find 
\st
\label{eqmostprob}
 \psi_{1,3}^{\rm mp} =  3\psi_{3,3} - 2\Psi_{PP} - \pi \, .
\stp
To estimate the degree of correlation between the 
most probable value and $\psi_{1,3}$
we calculate
\st
 -\dlangle \cos(\psi_{1,3} - 3\psi_{3,3} + 2\Psi_{PP} ) \drangle \, , 
\stp
and illustrate the result in \Fig{cospest}. 
We will use this correlation in \Sect{conclusion_a} to  make a definite
prediction for the behavior of two particle correlations 
with respect to  the reaction plane. 

\subsection{Convergence of the cumulant expansion for smooth Glauber type initial conditions}

\begin{figure}
\includegraphics[width=0.49\textwidth]{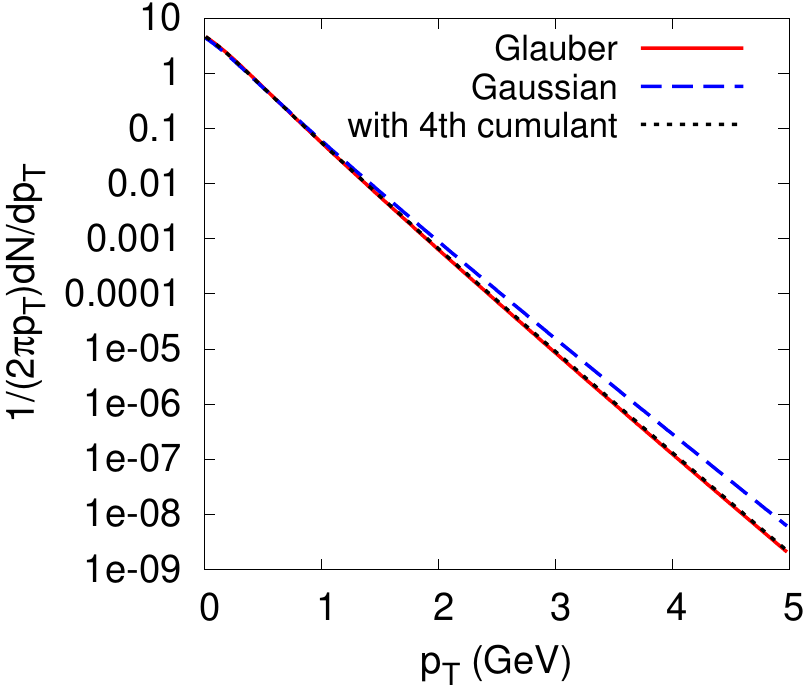}
\includegraphics[width=0.49\textwidth]{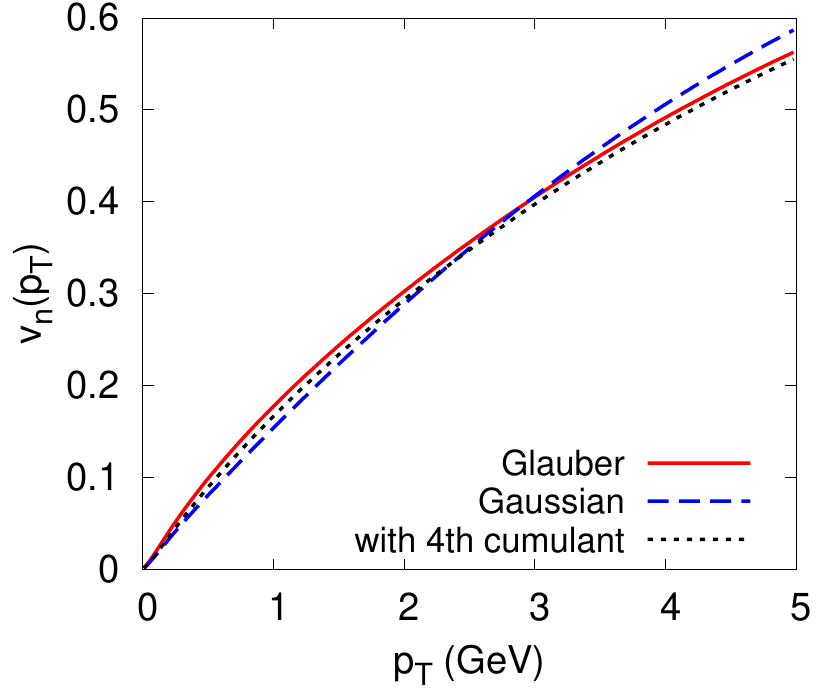}
\caption{
 (Color Online) (a) Spectra in the smooth (optical) Glauber  
model compared to the cumulant expansion.  The coefficients
of the Gaussian and fourth order cumulant expansions have been adjusted to 
reproduce $\llangle r^2 \rrangle$, $\llangle r^2\cos2\phi \rrangle$ 
and $\llangle r^4\cos2\phi \rrangle$,  $\llangle  r^4 \cos4\phi \rrangle$
respectively. The total entropy of the   cumulant
expansion  is also matched to the  total entropy of the glauber distribution.
(b) Elliptic flow in the Glauber model compared to the cumulant expansion. 
\label{cumulant_fig}
}
\end{figure}

In the previous section we introduced a cumulant expansion to characterize
the response of the system to a set of perturbations. In this section we 
will study the convergence of the cumulant expansion. Specifically, 
for a smooth (optical) Glauber profile, we will replace the initial
entropy distribution with an approximately Gaussian profile and 
cumulant corrections through  forth order.    
The distribution of entropy in the optical Glauber model (see \App{regulate}) is first 
used to calculate  $\llangle r^2 \rrangle$ and $\llangle r^2 \cos2\phi \rrangle$, which determines the two coefficients of the  Gaussian.  
Also the normalization ({\it i.e.} the total entropy) 
is the same between the Gaussian and the Glauber distribution. 
Taking the impact parameter to be $b=7.6\,$fm, 
\Fig{cumulant_fig} compares the spectra and the elliptic flow for these
two distributions. In the next approximation, the fourth cumulants 
to the Gaussian are adjusted as described in \Sect{dipole_tripole} and \App{regulate} to 
reproduce the $\llangle r^4 \rrangle$, $\llangle r^4 \cos 2\phi \rrangle$, 
and $\llangle r^4 \cos4\phi \rrangle$ moments of the Glauber distribution. 
\Fig{cumulant_fig} shows that the cumulant expansion reproduces the response
of the Glauber distribution in detail.

\section{Time development of the response } 
\label{time}

In the previous sections we introduced a set of initial
conditions with definite triangularity and
dipole asymmetry. In this section we will show how the 
hydrodynamic response to these cumulants develops in space and time. 
The point here is to understand the hydrodynamics without 
the complications of freezeout and a freezeout prescription. 

To show how the dipole and triangular flow develop in time, we have 
generalized the discussion of elliptic flow given in \Ref{Kolb:2000sd}.  
The spatial anisotropy is characterized by the
second moment 
\st
  \epsilon_{2x}  = - \frac{\llangle r^2 \cos 2\phi \rrangle}{\llangle r^2 \rrangle} \,  ,
\stp
which is  a function of time in general. As the system expands, the spatial 
anisotropy decreases and the momentum anisotropy increases. 
The momentum anisotropy is traditionally defined with $\epsilon_{2p}$ : 
\begin{align}
\label{epsilon2p_old}
  \epsilon_{2p} \equiv& \frac{\int \dd^2\x \, (T^{xx} -  T^{yy} )}{\int \dd^2\x \, \left(T^{xx}  + T^{yy} \right) }
 = \frac{\int \dd^2\x  \, (e + p) u_r^2 \cos2\phi_{u}  }{\int \dd^2\x\,   \left[(e + p) u_r^2 + 2p \right] } \, , 
\end{align}
where $u_r = \sqrt{(u^x)^2 + (u^y)^2}$ and $\phi_{u} = \tan^{-1} (u^y/u^x)$ .
This definition has its flaws  since the numerators and
denominators do not transform as components of a tensor under transverse 
boosts\footnote{This 
flaw is easily remedied by replacing  $\dd^2\x$ with 
the fluid three volume in the local rest frame
$d\Sigma_{\mu} u^{\mu} = \dd^2\x \dd\eta  \,  \tau u^{0}$.
The additional
factor of $u^{0}$ appears naturally below.}\, \cite{Teaney:2009qa}.
An alternative definition is  found by constructing an irreducible rank two tensor
% (with 
%respect to rotations around the beam axis) 
 out of the momentum density $T^{0i}$ 
and the flow velocity $u^{j}$
\st
  T^{0(i} u^{j)}  - \mbox{traces} \equiv  \frac{1}{2} \left( T^{0 i} u^{j} + T^{0j} u^i - \delta^{ij} T^{0l} u_{l} \right)  \, . 
\stp
Then we define
\begin{align}
 \epsilon_{2p} =&  \frac{\int \dd^2\x \tau  \left[ T^{0(x} u^{x) }  -\mbox{traces} \right]}{\int \dd^2\x \tau   \left[ T^{00} u^0  \right] } =  \frac{\int \dd^2\x \, \tau u^{0} \left[ (e + p) u_r^2 \cos2\phi_{u} \right] }{ \int \dd^2\x \, \tau u^{0} \left[ (e + p) u_r^2  + e \right]   }  \, , 
\end{align}
which is almost the same as \Eq{epsilon2p_old}. 
For the triangularity and dipole asymmetry we define the (reducible) third
rank tensor
\st
 T^{0(i} u^{j} u^{l) } =  \frac{1}{3!} \left(T^{0i} u^j u^l  + \mbox{perms} \right) \, . 
\stp
Then the traceless (or irreducible) tensor is used to define the momentum space triangular anisotropy 
\begin{align}
 \epsilon_{3p} \equiv&  \frac{\int \dd^2\x \tau  \left[ 
T^{0(x} u^{x} u^{x)} -\mbox{traces} \right]}{
\int \dd^2\x \tau   \left[ T^{00} u^0 u^0 \right] }  
 = 
\frac{\int \dd^2\x \,\tau u^{0} \left[ (e + p) u_r^3 \cos3\phi_u \right]  }
{\int \dd^2\x \tau  \left[ T^{00} u^{0} u^{0} \right]  } \, , 
\end{align}
and the trace is used to define momentum space dipole asymmetry
\begin{align}
 \epsilon_{1p} \equiv& \frac{\int \dd^2\x \tau \left[ \delta_{jl} T^{0(x} u^{j} u^{l)} \right] } {\int \dd^2\x\tau \left[ T^{00} u^0 u^0 \right]  } 
 = \frac{ \int \dd^2\x \,\tau u^{0} \left[ (e + p) u_r^3 \cos\phi_{u} \right] } {\int \dd^2\x\tau \left[ T^{00} u^0 u^0 \right]  } \, .
\end{align}

\begin{figure}
\begin{center}
\includegraphics[width=0.49\textwidth]{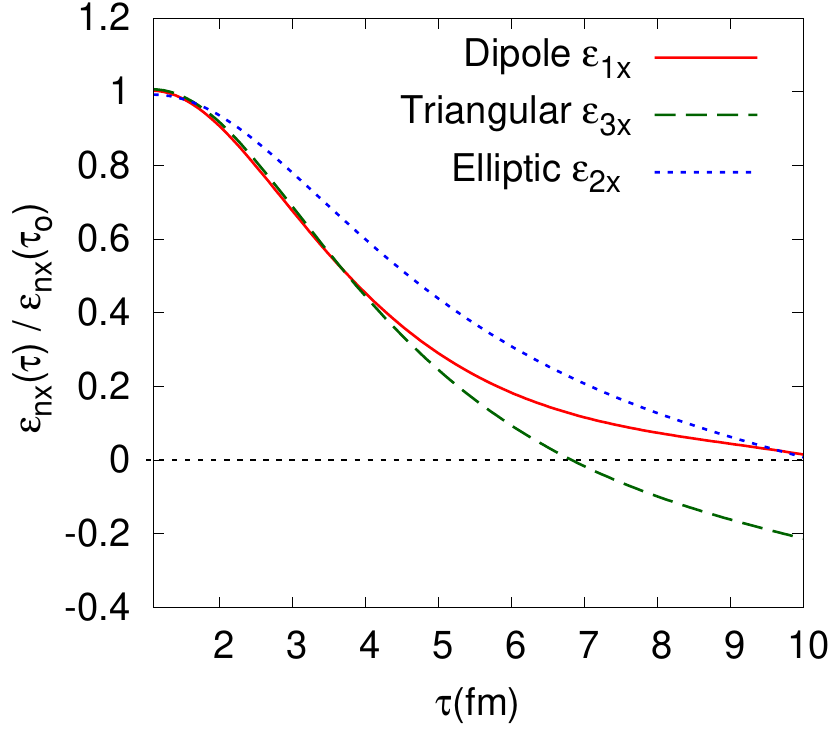}
\includegraphics[width=0.49\textwidth]{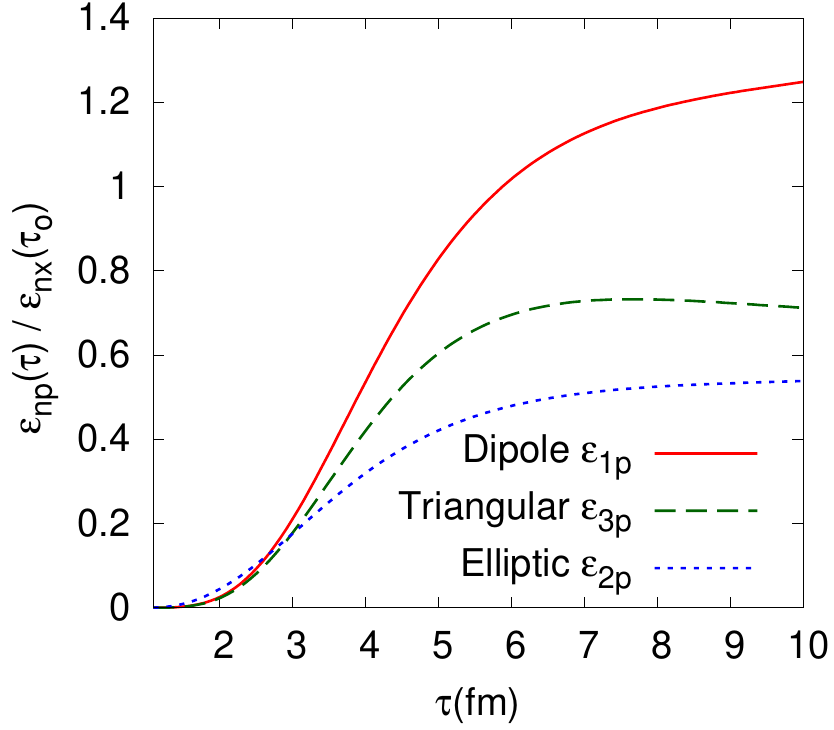}
\end{center}
\caption{(Color online) (a) The spatial anisotropy of the  entropy 
distribution $\epsilon_{1x}$, $\epsilon_{2x}$, and $\epsilon_{3x}$  (see text)
as a function of time for $b=7.6\,{\rm fm}$.  (b) The momentum anisotropy
$\epsilon_{1p}$, $\epsilon_{2p}$, and $\epsilon_{3p}$ (see text)  
as a function of time.  
The timescale in these figures
should be compared to $\sqrt{\llangle r^2 \rrangle}/c_s \simeq 5.4\,{\rm fm}$.
\label{epsilon_time} }  
\end{figure}

Armed with these definitions, \Fig{epsilon_time} illustrates the 
development of the triangular flow and the dipole asymmetry as
a function of time. As is familiar from studies of the elliptic 
flow \cite{Kolb:2000sd,Ollitrault:1992bk}, the spatial anisotropy decreases 
leading to 
a growth of the momentum space anisotropy. When the spatial
anisotropy crosses zero, the growth of the momentum space anisotropy
stalls. The figures also indicate that the elliptic flow, the dipole asymmetry, and the
triangularity all develop on approximately the same time scale,
 $\tau \simeq \sqrt{\llangle r^2 \rrangle}/c_s $.

\begin{figure}
\begin{center}
\includegraphics[width=0.49\textwidth]{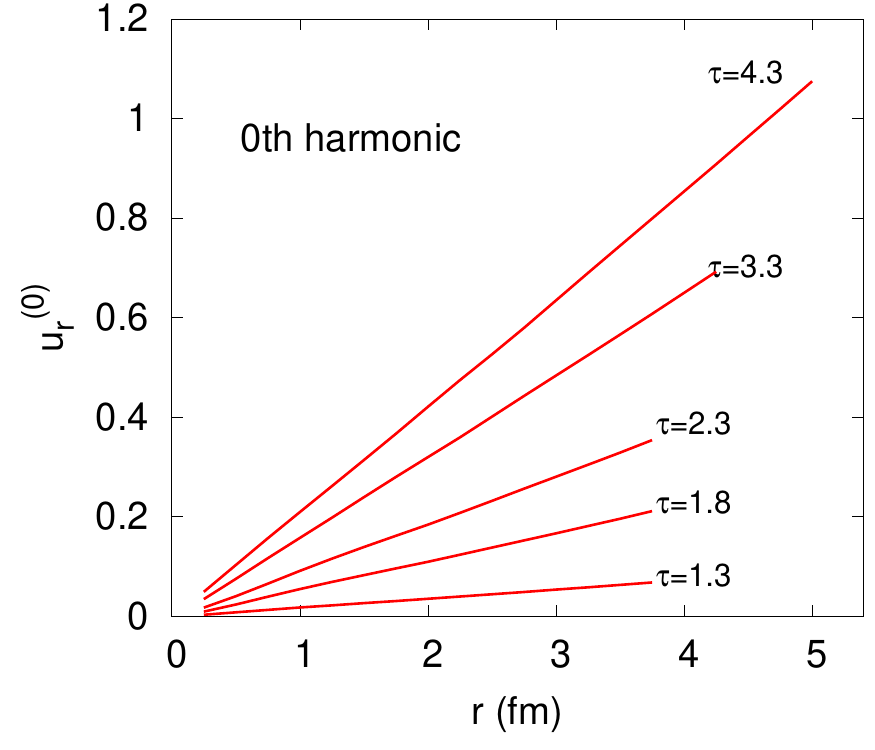}
\includegraphics[width=0.49\textwidth]{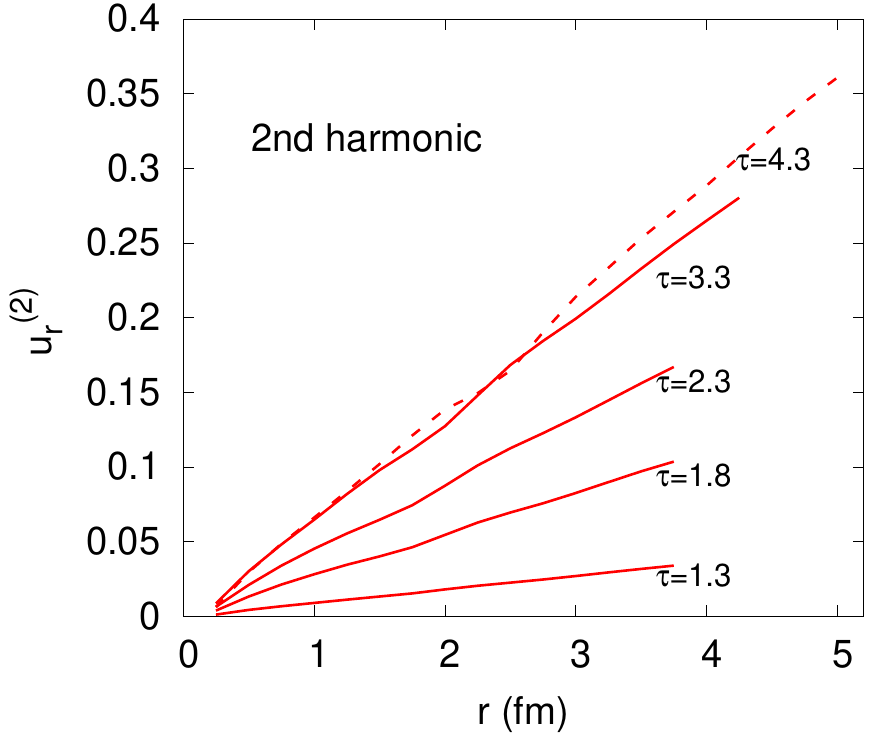}
\includegraphics[width=0.49\textwidth]{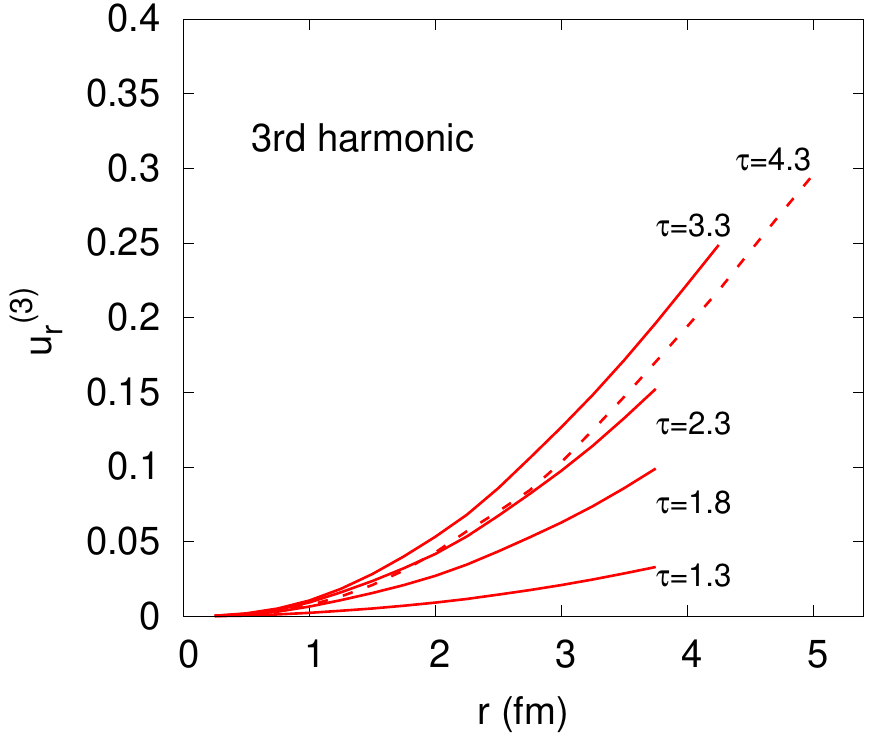}
\includegraphics[width=0.49\textwidth]{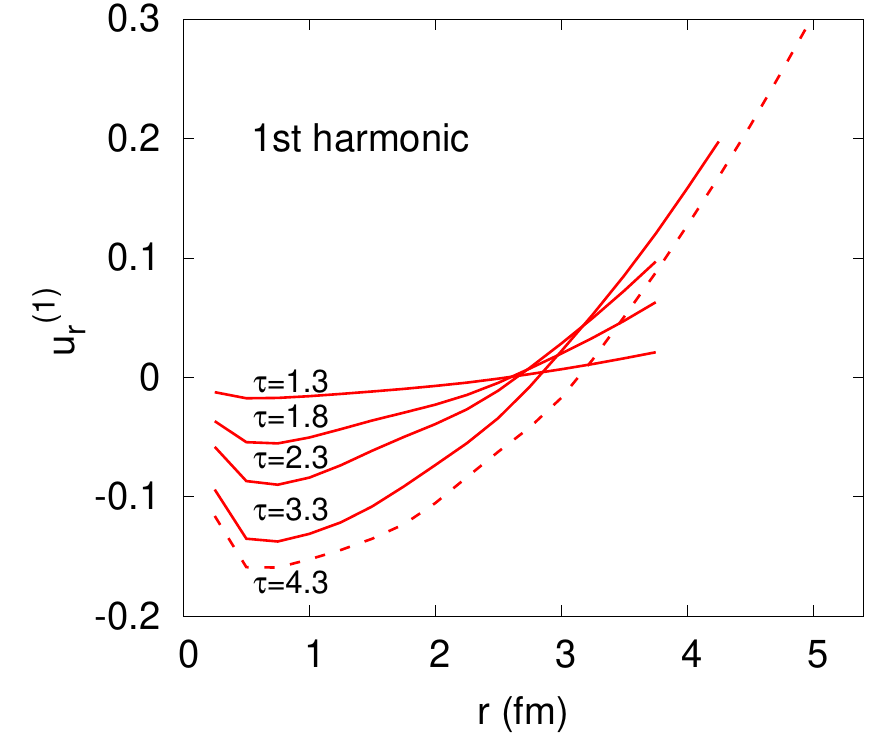}
\caption{(a) The zeroth harmonic of the flow profile (see Eq.~\protect\ref{transverse_flow}) 
for the radially symmetric Gaussian adopted in this work. The
root mean square radius of the Gaussian is adjusted to reproduce an 
impact parameter of $7.6\,{\rm fm}$.
(b) The second harmonic of the flow profile  for an elliptic perturbation.
(c) The third harmonic of the flow profile for a triangular perturbation
(d) The first harmonic of the flow profile for a distribution 
with a net dipole asymmetry. The deformations $\epsilon_1, \epsilon_2$ 
and $\epsilon_3$ are all set to $0.1$. 
\label{flow_profile}
}
\end{center}
\end{figure}

Another important aspect of the flow is the transverse radial flow profile.
To illustrate this profile we decompose the transverse flow velocity into
harmonics:
\begin{align}
\label{transverse_flow}
  u_{r}(r,\phi) =& u_{r}^0(r) + 2 u_{r}^{(1)}(r) \cos(\phi) + 2 u_{r}^{(2)}(r) \cos(2\phi) + 2 u_{r}^{(3) }(r) \cos(3\phi) + \ldots  \, .
\end{align}
For a radially symmetric Gaussian distribution only the zero-th harmonic
is present, and  $u_r^{(0)}$  shows a linearly rising
flow profile. When the elliptic deformation is added the second harmonic
also shows a linearly rising profile. Close to the origin this behavior 
can be understood with a linearized analysis of the acoustic waves.
The flow velocity in an acoustic analysis 
is the gradient of a scalar function $\Phi$  which can
be expanded in harmonics:
\st
  \Phi(r, \phi) = \Phi^{(0)}(r) +  2\Phi^{(2)}(r) \cos2\phi  + \ldots  \, .
\stp
If $\Phi(r,\phi)$ is an analytic function of $x$ and $y$, then 
$\Phi^{(2)}$ must be quadratic for small $r$.  Consequently the 
gradient of this function,   $u^{(2)}_{r}(r)$,  rises linearly at small $r$. 
Similarly, the  triangular deformation  $\Phi^{(3)}(r)$  should
be cubic at small $r$ and the flow profile $u^{(3)}_r$ should  be quadratic.
  These features are  borne out by our numerical work 
as exhibited in \Fig{flow_profile}. 
\Fig{flow_profile} also shows the flow profile of the 
first harmonic which results from an initial dipole asymmetry. 
The first harmonic shows
a  negative  slope at small $r$ followed by a quadratically
rising  profile at larger $r$.
   
As seen from \Fig{flow_profile}, the triangular and dipolar flows are
biased towards the edge of the nucleus.
In the next section we will see that 
due to this bias
$v_{1}$ and $v_3$
are more sensitive to the freezeout prescription than  $v_2$.

\section{Particle spectra: $v_1(p_T)$ and $v_3(p_T)$ }
\label{spectra}

Having illustrated the essential features of the hydrodynamic response,
we will compute the particle spectra associated with these flows. 
As discussed above, the analysis is limited to a classical massless
ideal gas.  We will follow the time honored, but poorly motivated 
prescription of specifying a freezeout temperature or a freezeout entropy density. 
Freezeout temperatures in full hydrodynamic simulations with a Hadronic
Resonance Gas  (HRG) range from $T=160\, {\rm MeV}$ to $T=120\,{\rm MeV}$ \cite{Kolb:2003dz,Teaney:2009qa}.  
The total initial entropy and initial volume used in our massless ideal gas 
simulations  were taken to be the similar to the total entropy  
and initial volume used in these full hydrodynamic 
simulations.  The final freezeout volume of our massless-gas simulation is
also taken to be similar to the final freezeout volume of these full 
simulations. Since entropy is conserved, this can be accomplished
by adjusting the freezeout entropy density of the 
massless gas so that the entropy density equals the  HRG entropy density for 
a specified HRG freezeout temperature. 
Experience has shown that this is a fair way to compare different equations of state.
Rather than quoting the actual freezeout temperature of the massless
gas EOS, we will simply quote the corresponding HRG freezeout temperature.
Thus $T\Leftrightarrow 170\,{\rm MeV}$ means that the actual freezeout 
temperature is such that the entropy density of a massless gas is
equal to the entropy density of the HRG at $T=170\, {\rm MeV}$.
Table~\ref{FOTable} 
shows a set of temperatures and entropy densities in a HRG model 
and the corresponding freezeout temperatures for the massless gas equation of state.
\begin{table}
\begin{tabular}{c|c|c|}
Hadron Gas $T_{\rm fo}$ &  Hadron Gas $s_{\rm fo}$  &  Massless Gas  $T_{\rm fo}$ \\ \hline
$130\, {\rm MeV}$  & $4.34\, {\rm fm}^{-3}$  & 71 \, {\rm MeV}  \\
$150\, {\rm MeV}$  & $1.87\, {\rm fm}^{-3}$  & 96 \, {\rm MeV}   \\
$170\, {\rm MeV}$  & $0.77\, {\rm fm}^{-3}$  & 127\, {\rm MeV}  \\
\end{tabular}
\caption{Table of freezeout temperatures used in this work.
The first two columns show freezeout temperatures and the corresponding
entropy densities of a Hadron Resonance Gas (HRG) EOS.
The last column shows the freezeout temperatures 
where the massless gas EOS used in this work attains the corresponding
HRG entropy density.
\label{FOTable}
}
\end{table}

\begin{figure}[h!]
\includegraphics[height=0.5\textwidth]{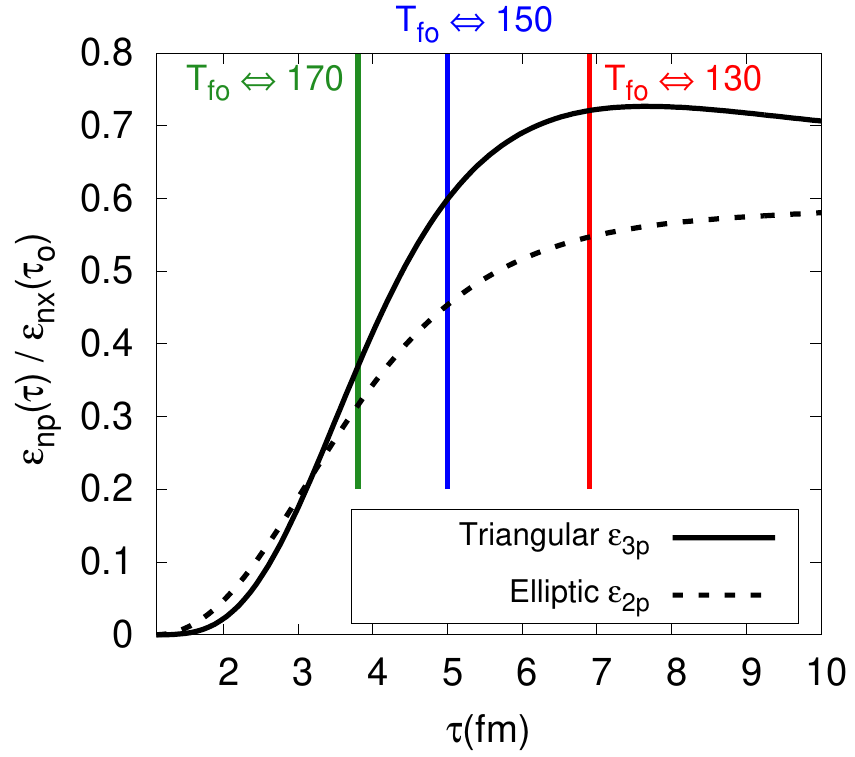}
\caption{Evolution of the momentum anisotropy as a function 
of time at an impact parameter of $b=7.6\,{\rm fm}$. The lines
indicate when the average entropy density $\llangle s \rrangle$
 falls below the freezeout entropy density specified by the 
the temperatures $T \Leftrightarrow 130,150,170\, {\rm MeV}$.  
\label{FOTimes}}
\end{figure}

\Fig{FOTimes} 
shows the momentum anisotropies as a function of time, and 
marks when the average entropy density of the system reaches a specified freezeout entropy
density. Specifically, the lines indicate when 
$\llangle s \rrangle$  in the notation of \Eq{avedef}  falls
below the freezeout entropy density indicated in Table ~\ref{FOTable}. 
We see that for $T_{\rm fo} \Leftrightarrow 170\, {\rm MeV}$ the 
triangular and dipole flows are still developing, 
while for $T_{\rm fo} \Leftrightarrow 130\,{\rm MeV}$ 
the flows are almost fully developed.

\begin{figure}
\begin{center}
\includegraphics[width=0.49\textwidth] {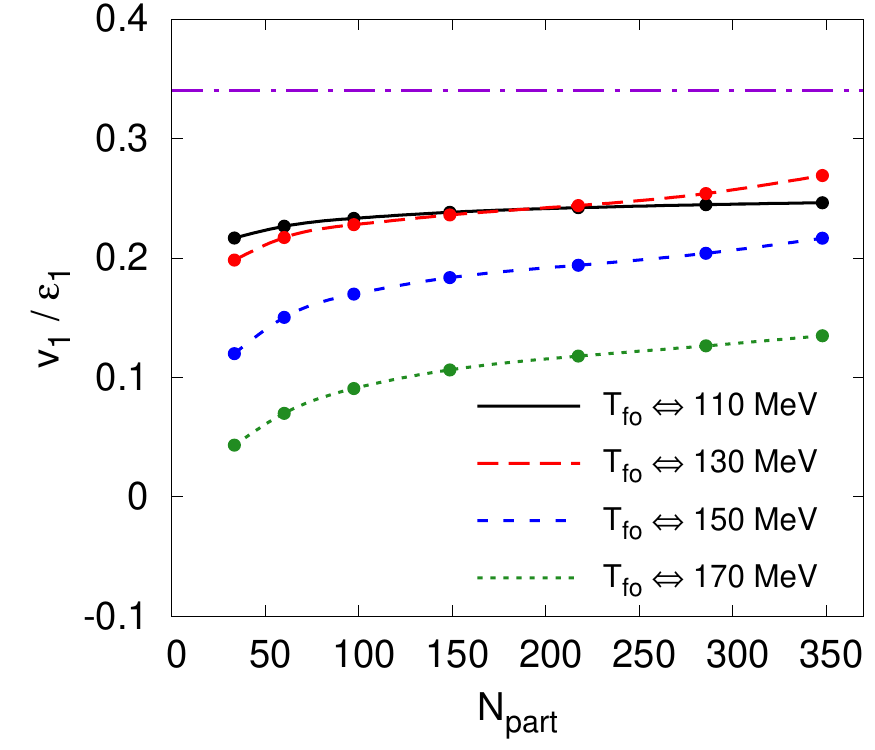}
\includegraphics[width=0.49\textwidth  ]{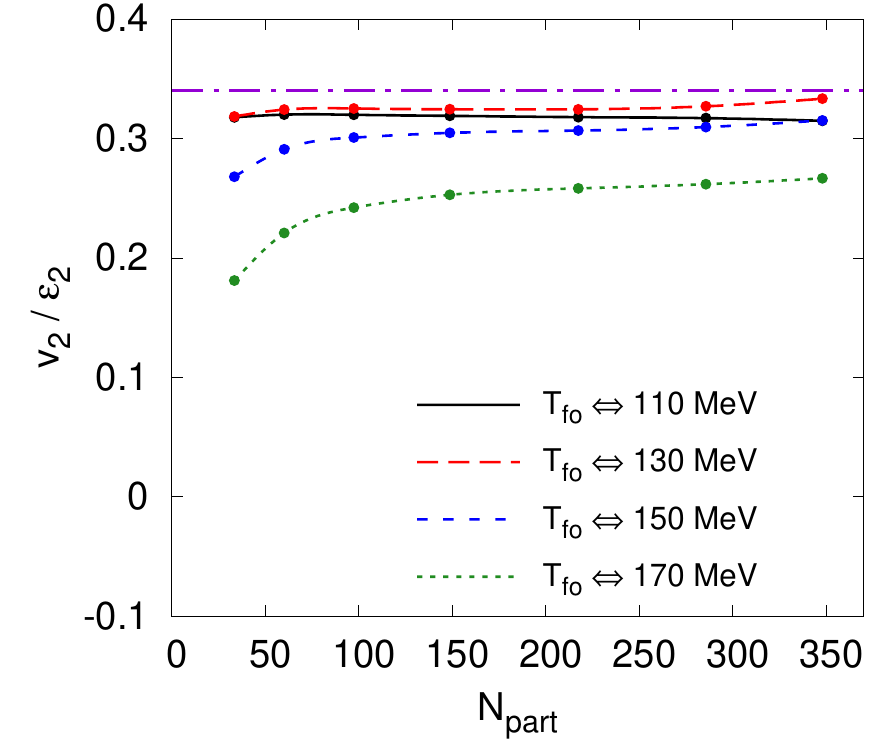}
\includegraphics[ width=0.49\textwidth ]{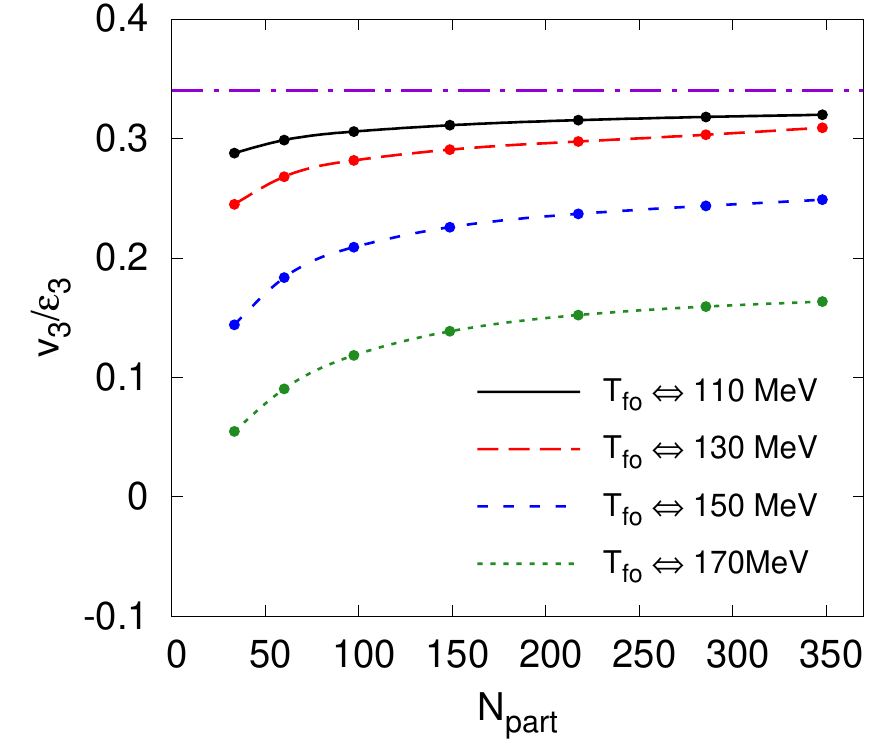}
\end{center}
\caption{$v_1$, $v_2$ and $v_3$ per unit anisotropy as 
a function of $N_{\rm part}$ for different freezeout temperatures.  The anisotropy
parameters are all $0.1$ in the actual simulations. 
\label{vnvsnpart}
}
\end{figure}

Once the freezeout surface is specified the particle spectra 
are computed using the Cooper-Frye formula
\st
   (2\pi)^3 E \frac{dN}{d^3\p} = \int_{V} p^{\mu} dV_{\mu}  \, f_o(-P \cdot U(X)) \, , 
\stp
where  $f_o(E) = g\exp(-E/T(X)) $ is the distribution
function of a classical massless gas. (The notation here 
follows the review article \cite{Teaney:2009qa}.)  Using this formula
we compute the particle spectra and determine 
the associated harmonics $v_1$, $v_2$ and $v_3$.  For each impact parameter
we determine the root-mean square radius and the total entropy from an 
optical Glauber model;
then the Gaussian parameters are adjusted to reproduce these Glauber quantities; 
finally the simulation is run to the freezeout entropy density and 
the harmonics are computed.  \Fig{vnvsnpart} shows how the 
harmonics depend on centrality and the freezeout temperature.

Examining \Fig{vnvsnpart} we see that $v_1$, $v_2$ and $v_3$
are roughly independent of centrality. However, it must be 
borne in mind that in a more complete simulation, the total
entropy per participant is  also a function of centrality and this
could change the result. Here
the entropy per participant is constant. 
Generally
the freezeout criterion is also a function of centrality and
this could give a substantial shape to these curves in a final 
simulation. Finally, when viscosity is included the triangularity
is also a more complicated function of centrality \cite{Alver:2010dn}. This
will be explored elsewhere \cite{YanWork}.

\begin{figure}
\begin{center}
\includegraphics[ width=0.48\textwidth ]{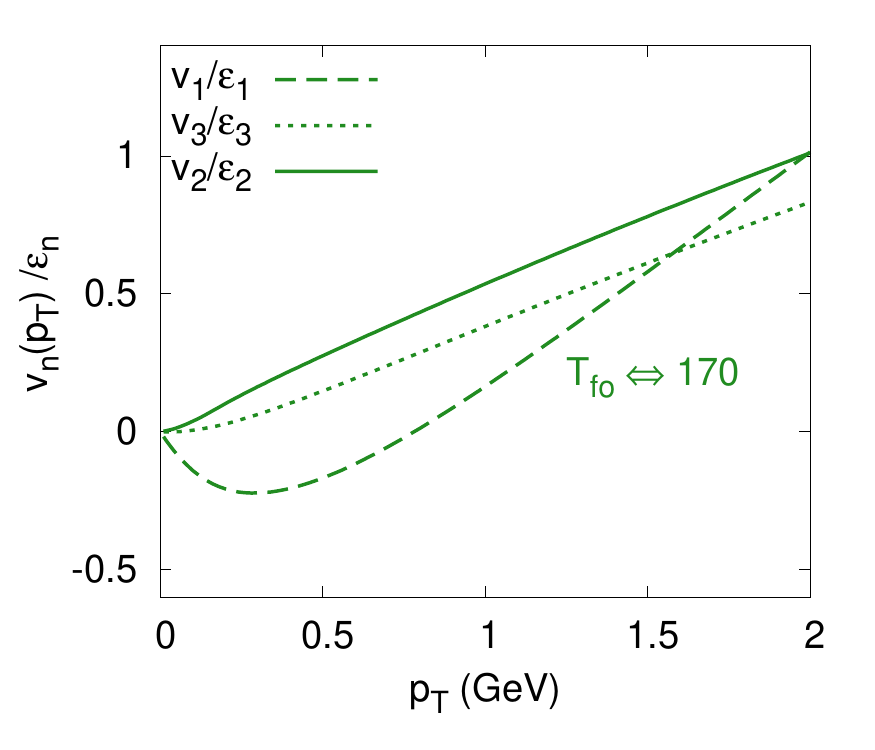}
\includegraphics[width=0.48\textwidth  ]{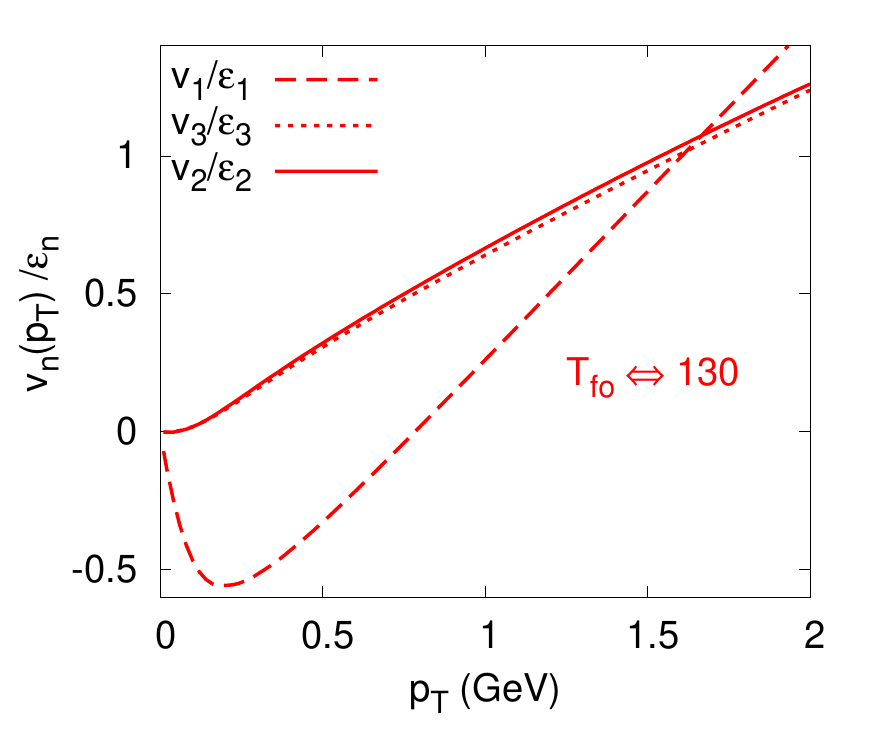}
\end{center}
\caption{$v_n(p_T)$ for two different freezeout temperatures
as described in Table~\protect\ref{FOTable}.  The root mean square
radius of the initial Gaussian corresponds to a radius of $b=7.6\,{\rm fm}$ 
\label{vnofpt}}
\end{figure}

\Fig{vnofpt} shows how these harmonics depend on $p_T$.  $v_2(p_T)$
and $v_3(p_T)$ show a characteristic linear rise with $p_T$ that
is a consequence of a strong radial 
flow \cite{Huovinen:2001cy, Borghini:2005kd,Mishra:2007tw, Mishra:2008dm}.  
Indeed examining the thermal distribution with constant temperature, we have 
\begin{align}
 e^{P\cdot U/T} =& e^{-E_\p U^{\tau}/T}e^{p_{T}/T\, u_{r}(r,\phi) \cos(\phi_p - \phi_{u}) } \, ,  \\
 \simeq &  e^{-E_\p /T}e^{p_{T}/T u_{r}^{(0)} (r) \cos(\phi_p - \phi) } \nonumber \\
  & \quad\times \left[1 +   \frac{2p_{T}}{T} u_r^{(2)}(r)\cos(2\phi) \cos(\phi_\p - \phi) +
\frac{2p_{T}}{T} u_r^{(3)}(r)  \cos(3\phi) \cos(\phi_\p - \phi) + \ldots \right] \,  .
\end{align}
Here $E_{\p}$ is the energy, $\phi_{\p}$ is the particles azimuthal angle;
  we have adopted a non-relativistic approximation $U^{\tau} \simeq 1$ and assumed that the flow is approximately radial, $\phi_{u} \simeq \phi$.  Further,
we have neglected $u_{r}^{(1)}$ in this discussion.
Unless the momentum angle equals the spatial angle $\phi_\p \simeq \phi$, the thermal distribution is strongly suppressed by 
the leading Boltzmann factor. Thus, we
arrive at a form which illustrates the linear rise of  rise of $v_n(p_T)$ with $p_T$
\begin{align}
 e^{P\cdot U/T} 
 \simeq &  e^{-E_\p /T}e^{p_{T}/T u_{r}^{(0)} (r)  }  \left[1 +   \frac{2p_{T}}{T} u_r^{(2)}(r) \cos(2\phi_\p)  + \frac{2p_{T}}{T} u_r^{(3)}(r)  \cos(3\phi_\p)  + \ldots \right] \,  .
\end{align}
Examining \Fig{vnofpt}, we see that $v_1(p_T)$ also
displays a similar linearly rising trend at higher $p_T$ after an initial dip.

\section{Further predictions and comparison with other works. }

\subsection{Further predictions}
\label{conclusion_a}
\Fig{vnvsnpart} and \Fig{vnofpt} show the response of the 
hydrodynamic system to the deformations. Certainly it is premature
to compare the current calculation to data. For instance, the
effect of viscosity, resonance decays, and a lattice-based equation
of state have not been included. These reality factors will 
reduce the response. Nevertheless, in order to keep the final
goal clearly in sight, we will provisionally compare the current
calculation to the Alver Roland fit \cite{Alver:2010gr} of STAR inclusive 
two particle correlations \cite{Abelev:2008un}. Further, we will suggest a number
of additional observables which can confirm the geometric 
nature of the measured two particle correlations. 

The average over glauber configurations 
at fixed $N_{\rm part}$ is denoted with 
double brackets $\dlangle\ldots \drangle$. 
Then the 
two particle angular correlation function 
%of a set particle labeled by $\alpha$ and
%set of particles labeled by $\beta$ at fixed $N_{\rm part}$
can be expanded in a Fourier 
series:
\begin{align}
\label{corrparam}
\Dlangle \frac{\dd N_{\rm pairs,\alpha\beta}}{\dd \phi_\alpha \dd \phi_\beta} \Drangle 
= \dlangle N_{\rm pairs,\alpha\beta}  \drangle \left(1  + \sum_{n} 2 V_{n\Delta } \cos(n\phi_\alpha - n\phi_\beta)  \right) \, .
\end{align}
The particle labels $\alpha$ and $\beta$ could denote  distinct particle types or  $p_T$ bins for example.
Following  Alver and Roland we will 
approximate the two particle correlation with the disconnected component.
The yield of particle type $\alpha$ for a fixed Glauber configuration is
\begin{multline}
\label{eq:response}
\frac{\dd N_\alpha}{\dd \phi_\alpha} =  
\frac{N_{\alpha}}{2\pi} \Big[1 
+ 2 \frac{v_{1\alpha}}{\epsilon_1} \epsilon_1 \cos(\phi_{\alpha} - \psi_{1,3}) 
+ 2 \frac{v_{2\alpha}}{\epsilon_2} \epsilon_2 \cos(2\phi_\alpha - 2\Psi_{PP}) \\
+ 2 \frac{v_{3\alpha}}{\epsilon_3}  \epsilon_3 \cos(3\phi_\alpha - 3\psi_{3,3})
\Big] \, ,
\end{multline}
where we have assumed that the response  is linearly proportional to
the deformation.
Then the two particle correlation function is approximated as
\begin{multline}
\label{integrated_pred}
\Dlangle \frac{\dd N_{\rm pairs,\alpha\beta}}{\dd\phi_\alpha \dd\phi_\beta} \Drangle 
 \simeq   
\Dlangle \frac{\dd N}{\dd\phi_\alpha}  \frac{\dd N}{\dd \phi_\beta} \Drangle 
\simeq  \frac{N_\alpha N_\beta}{(2\pi)^2} \left[ 
1  + \sum_{n} 2 \left( 
\frac{v_{n\alpha} v_{n\beta} }{\epsilon_n^2} 
\right)  
\dlangle \epsilon_n^2  \drangle
\cos(n (\phi_\alpha - \phi_\beta))   \right]  \, .
\end{multline}
Here and below we have 
tacitly assumed that the multiplicity fluctuations at
fixed $N_{\rm part}$ are negligible.  If this is not the case 
then one  has the following replacements in \Eq{integrated_pred} 
\st
  N_{\alpha} N_{\beta} \rightarrow \dlangle N_\alpha N_\beta \drangle 
 \qquad  \dlangle \epsilon_n^2 \drangle \rightarrow \frac{\dlangle N_\alpha N_\beta \epsilon_n^2 \drangle   }{\dlangle N_\alpha  N_\beta  \drangle } \, .
\stp

\begin{figure}
\begin{center}
\includegraphics[width=0.49\textwidth]{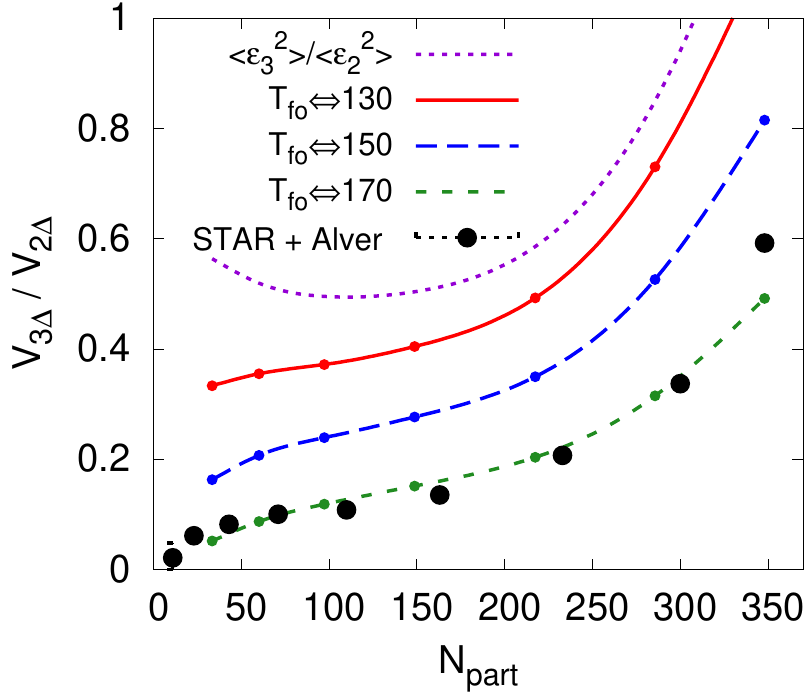}
\includegraphics[width=0.475\textwidth]{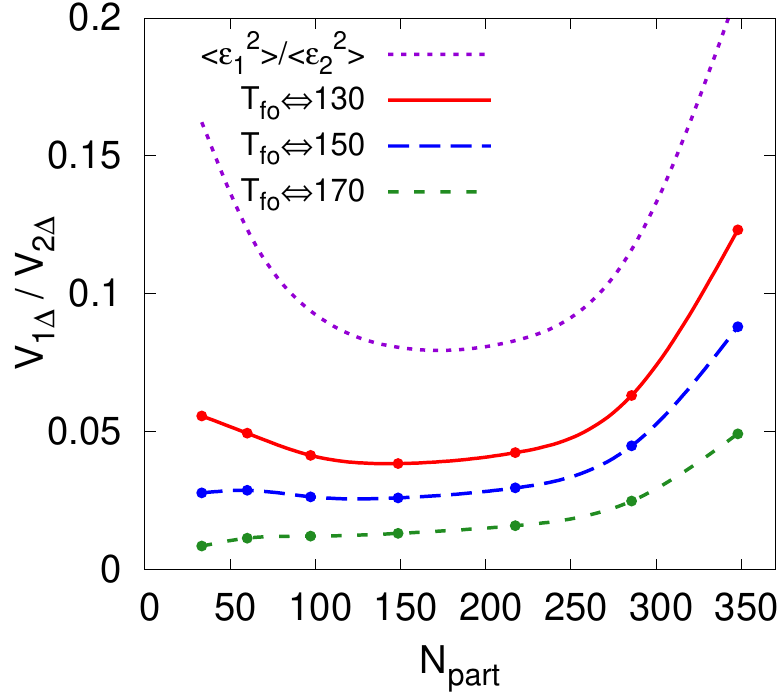}
\end{center}
\caption{ 
Fourier components of the two particle correlation function 
as a function of $N_{\rm part}$ relative to the quadrapole
component. (a) The triangularity component compared to 
the Alver Roland fit \protect \cite{Alver:2010gr} of STAR inclusive two particle
correlation functions \protect \cite{Abelev:2008un}. (b) The dipole component
relative to the quadrapole component; note that the scales differ between(a) and(b).
\label{twoparticlecorr}
}
\end{figure}

Given the parameterizations in \Eq{corrparam} and \Eq{integrated_pred}, the response functions
in \Fig{vnvsnpart}  make a definite prediction for
the different Fourier components $V_{n\Delta}$. The elliptic flow is too 
large in the ideal massless gas model considered here.  
We will therefore simply plot the ratios of the different fourier
components as was done in the Alver and Roland paper. 
Using the response functions in \Fig{vnvsnpart}, and the Glauber 
estimates for $\dlangle \epsilon_3^2 \drangle /\dlangle \epsilon_2^2 \drangle$,
\Fig{twoparticlecorr}(a) compares the strength of the triangular 
component  to the quadrapole component.  
The ideal hydrodynamic prediction (with a massless ideal gas EOS) 
is generally too large and fairly sensitive
to the freezeout temperature. This sensitivity reflects the fact
that the triangular flow develops further towards the edge of the nucleus.
\Fig{twoparticlecorr}(b) compares the dipole
component to the quadrapole component. The dipole
component is a factor of eight smaller than the  quadrapole component. 
This is a reflection of the fact that $\epsilon_1$ is  small, 
and the fact that $v_1(p_T)/\epsilon_1$ is  positive and negative.
The dipolar flow is also sensitive to the details of freezeout.

Next we wish to determine the general form of the two particle
correlation function with respect to the participant plane $\Psi_{PP}$ 
\st
\label{twopart_wrtR}
\Dlangle \frac{\dd N_{\rm pairs,\alpha\beta}}{\dd \phi_1 \dd \phi_2} \Drangle_{\Psi_{PP} }
\simeq  \Dlangle
\frac{\dd N_{\alpha}}{\dd(\phi_\alpha-\Psi_{PP})} \frac{\dd N_\beta}{\dd(\phi_\beta-\Psi_{PP})} 
\Drangle_{\Psi_{PP}} \, .  \\
\stp
Inserting \Eq{eq:response}  into \Eq{twopart_wrtR} and 
averaging over glauber configurations several  several terms appear. 
In \Sect{glauber} we identified
the principle correlations that exist between 
the angles $\psi_{1,3},\psi_{3,3}$ and $\Psi_{PP}$. Namely, 
the only significant fourier expectation values are 
$\llangle \cos(2\psi_{1,3} - 2\Psi_{PP}) \rrangle$ 
(as determined
by the coefficient $A$ in \Eq{Acoeffdef}), and 
$\llangle \cos(\psi_{1,3} - 3\psi_{3,3} + 2\Psi_{PP}) \rrangle$ (as determined
by the coefficient $B_0$ in \Eq{B0coeffdef}).   With the assumption 
that these are the only significant fourier expectation values
at third order, the form of the 
two particle correlation function with respect to participant plane becomes:
\begin{align}
\Dlangle \frac{\dd N_{\rm pairs,\alpha\beta}}{\dd\phi_\alpha \dd\phi_\beta} \Drangle  
\simeq
 \frac{N_{\alpha} N_{\beta}} {(2\pi)^2 } 
  \Big[ 
1  +& \sum_{n} 2 \left( 
\frac{v_{n\alpha} v_{n\beta} }{\epsilon_n^2} 
\right)  
\dlangle \epsilon_n^2  \drangle
\cos(n \phi_\alpha - n\phi_\beta)   \nonumber \\ 
+& \, \,
2\frac{v_{2\alpha}}{\epsilon_2} \dlangle \epsilon_2 \drangle 
 \; \cos(2\phi_\alpha - 2\Psi_{PP})   \nonumber \\
+& \, 
\, 2\frac{v_{2\alpha} v_{2\beta} }{\epsilon_2^2} \dlangle \epsilon_2^2 \drangle  \; \cos(2\phi_\alpha + 2\phi_\beta - 4\Psi_{PP} ) \nonumber \\
+& \, \, 
\, 2\frac{v_{1\alpha} v_{1\beta} }{\epsilon_1^2} \dlangle \epsilon_1^2 \cos(2\psi_{1,3} - 2\Psi_{PP})  \drangle 
 \; \cos(\phi_\alpha + \phi_\beta - 2\Psi_{PP} ) \nonumber \\
+&
\,
\, 2\frac{v_{1\alpha} v_{3\beta} }{\epsilon_1\epsilon_3} \dlangle \epsilon_1\epsilon_3 \cos(\psi_{1,3} - 3\psi_{3,3} + 2\Psi_{PP})  \drangle 
\cos(\phi_\alpha - 3\phi_\beta + 2\Psi_{PP} )  \nonumber \\
+&\,\,  \alpha \leftrightarrow \beta  \Big]  \, . 
\end{align}
The symmetrization 
with respect to $\alpha$ and $\beta$ 
applies to all terms in this expression which are not already symmetric, {\it e.g.} $\cos(2\phi_\alpha - 2\Psi_{PP})$. We will discuss this expression 
line by line.  
The first three lines are not particularly novel:
The first line is independent of the reaction plane angle $\Psi_{PP}$.
The next two lines reflect the underlying elliptic flow and 
would normally be subtracted in a flow subtracted two particle 
correlation  function. 

\begin{figure}
\begin{center}
\includegraphics[height=2.92in]{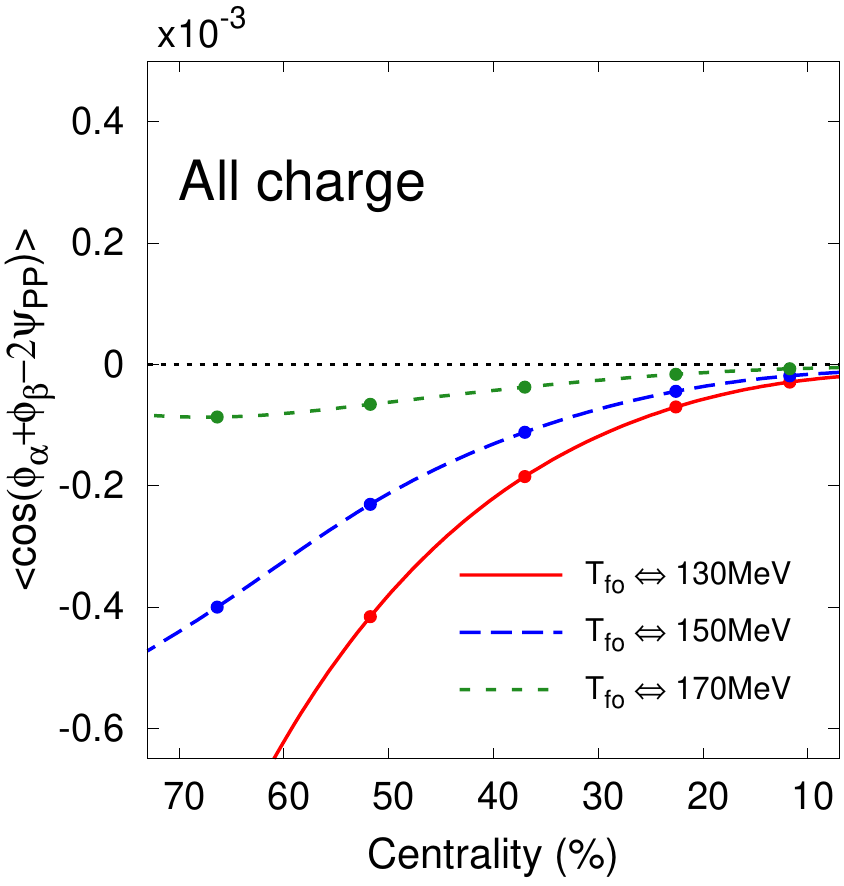} 
\includegraphics[height=3.04in]{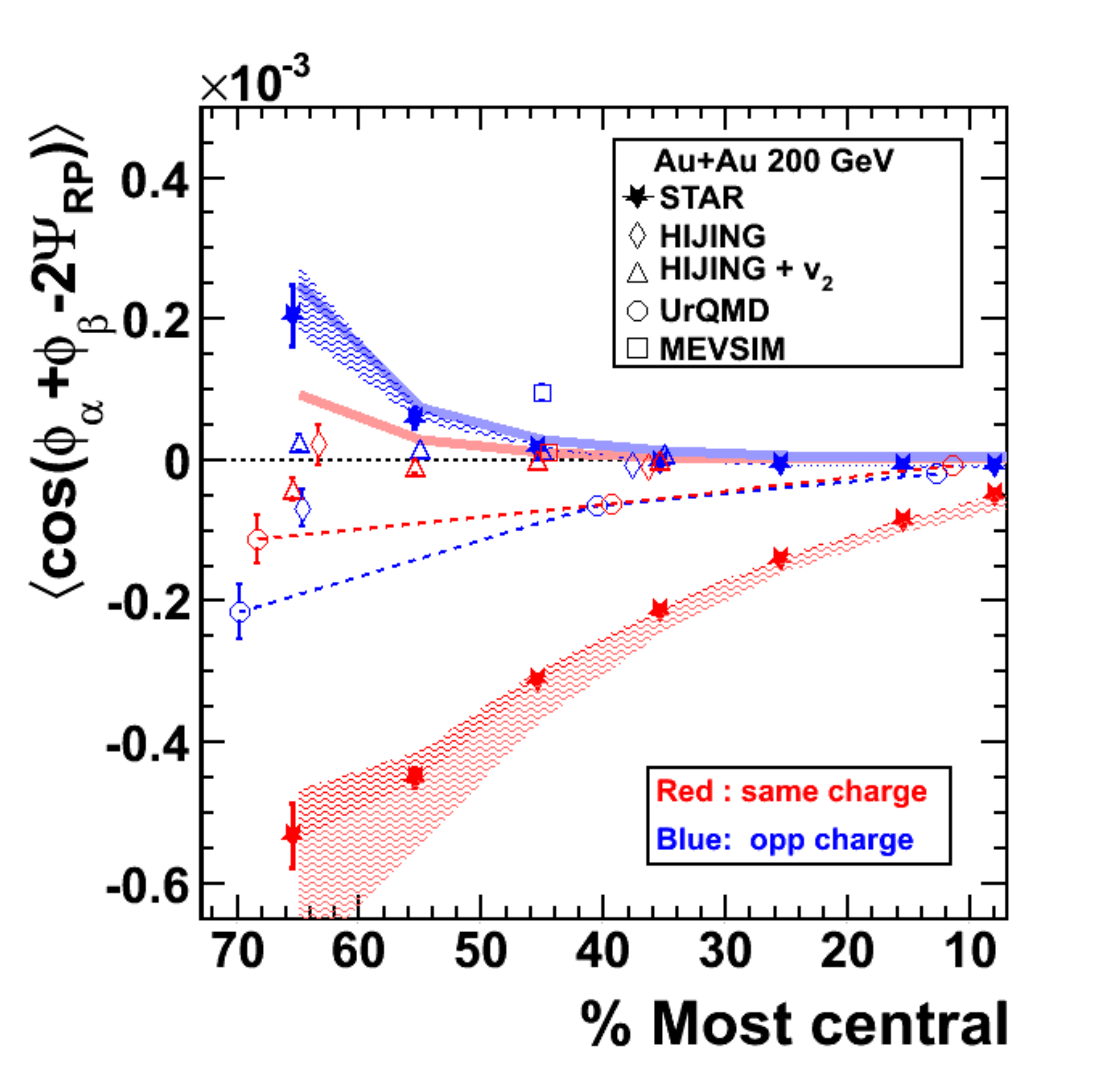}
\end{center}
\caption{
(a) The expectation value $\llangle \cos(\phi_\alpha + \phi_\beta - 2 \Psi_{PP} ) \rrangle$
as  predicted by hydrodynamics,  where
$\alpha$ and $\beta$ label all particles. (b) The charge asymmetry with respect to 
reaction plane $\llangle \cos(\phi_{\alpha} +  \phi_{\beta} - 2\Psi_{R}) \rrangle$ 
as measured by the STAR collaboration \protect \cite{Abelev:2009uh,Abelev:2009txa}. Here $\alpha$ and $\beta$ label $++,+-,$ or $--$.
%In this 
%case $\alpha$ and $\beta$ label specific charged species. 
The hydrodynamic prediction does not explain the charge asymmetry.
\label{dimaeffect}
}
\end{figure}

The fourth line contains the  first novel feature. This term
arises because the dipole asymmetry  is   
preferentially oriented out plane, leading to a net $v_1$ out of plane. 
\Fig{dimaeffect}(a) shows the correlation function $\llangle \cos(\phi_\alpha +
\phi_\beta - 2\Psi_{PP}) \rrangle$ as a function 
of centrality. Recently, the  STAR collaboration measured  a
similar  expectation value, but divided correlation function 
into the different possible charge components ($i.e.$++, +{-}, {-}{-}) 
in order to investigate the possibility of local parity violation in 
heavy ion collisions \cite{Abelev:2009uh,Abelev:2009txa,Kharzeev:1998kz}.  \Fig{dimaeffect}(b) shows the 
measured STAR correlations.  The measured correlation is the same 
order of magnitude as the out of plane flow found in this
work. However many 
aspects of the out of plane dipole flow (e.g. the $p_T$ dependence and 
most importantly the charge dependence) do not agree with the measured correlation.
Thus the STAR measurements can constraint the geometric fluctuations reported 
here. This will be investigated in future work.

\begin{figure}
\begin{center}
\includegraphics[width=0.5\textwidth]{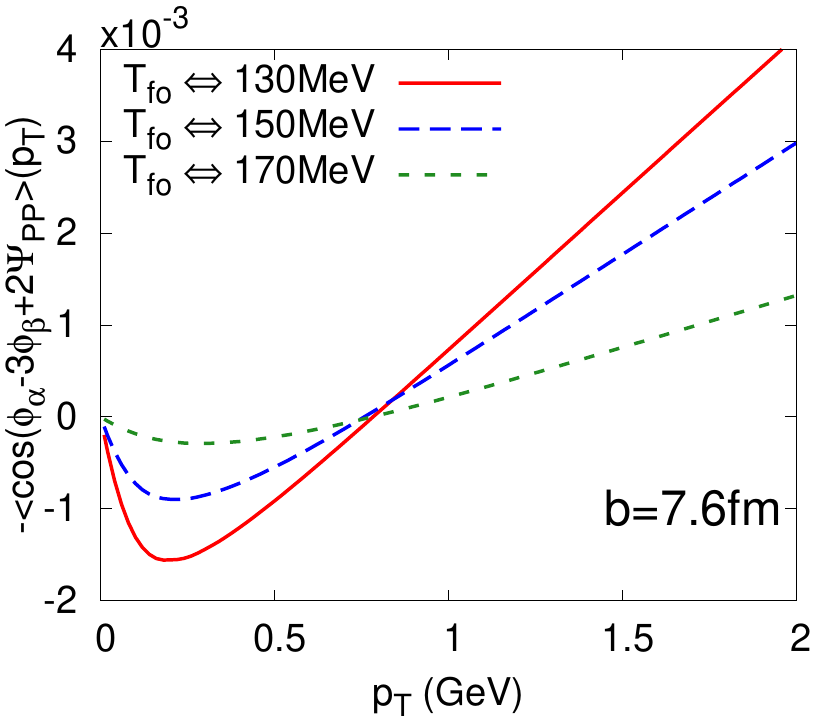}
\end{center}
\caption{  
A hydrodynamic prediction for the expectation value 
$\llangle \cos(\phi_\alpha- 3\phi_\beta + 2\Psi_{PP} ) \rrangle$ 
which reflects the correlation 
between the dipole, triangular, and elliptic event planes. 
Here $\alpha$ labels all particles in a given $p_T$ bin and $\beta$ 
labels all particles.
\label{coolplot}
}
\end{figure}

A second novel feature expressed by the two particle correlation function
with respect to reaction plane is recorded by the 5th line of \Eq{twopart_wrtR}.
It shows that hydrodynamics, together with the geometric fluctuations
of the Glauber model makes a definite prediction  for the angular correlation
\st
  \dlangle  \cos(\phi_{\alpha} - 3 \phi_{\beta} + 2\Psi_{PP} ) \drangle  \, .
\stp 
Taking $\alpha$ to label all the particles in a definite $p_T$ bin
and $\beta$ all the particles, this definite prediction reads
\st
  \dlangle  \cos(\phi_{\alpha} - 3 \phi_{\beta} + 2\Psi_{PP} ) \drangle   
 = \frac{v_1(p_T) }{\epsilon_1 } \frac{v_{3}}{\epsilon_3}  
\dlangle \epsilon_1 \epsilon_3 \cos(\psi_{1,3} - 3 \psi_{3,3}  + 2 \Psi_{PP}) \drangle \, .
\stp 
This result is illustrated in \Fig{coolplot} and 
is based on the Glauber analysis in \Fig{cospest} and the response functions
calculated in \Fig{vnofpt}.  Another way to probe this same correlation is
the following. Experimentally, the participant plane $\Psi_{PP}$ is traditionally estimated
 by using the standard $Q$ vector method, or the Yang-Lee zero
generalization  of this idea \cite{Voloshin:2008dg}. These same methods can be used to
determine the triangularity event plane $\psi_{3,3}$ without significant
modifications \cite{Voloshin_private}.  The strong correlation between the
dipole, triangular, and participant planes implies
that the knowledge of $\psi_{3,3}$ and $\Psi_{PP}$ determines the dipole
event plane $\psi_{1,3}$ at least statistically. The most probable orientation
is given by \Eq{eqmostprob}  and is repeated here for convenience
\[
 \psi_{1,3}^{\rm mp} =  3\psi_{3,3} - 2\Psi_{PP} - \pi \, .
\]
Thus, the $v_1$ associated with the dipole 
asymmetry can be determined  by measuring the expectation value
\st
  \dlangle \cos(\phi - \psi_{1,3}^{\rm mp} ) \drangle \, . 
\stp  
Essentially this correlation is  
a $v_1$ with an extra twist to take out the shifting 
orientations of the dipole and triangular event planes -- see  \Fig{corrfig}.

\subsection{Discussion and comparison with other works }
\label{conclusion_b}

We hope that the cumulant expansion presented in \Sect{cumulant} organizes  and formalizes  
the study of fluctuations in heavy ion collisions. 
The convergence of the cumulant expansion is really quite good as
illustrated in \Fig{cumulant_fig}.
At third order in the cumulant expansion there are two 
additional terms, the triangularity $\llangle r^3 \cos3(\phi - \psi_{3,3}) \rrangle $, and the dipole asymmetry $\llangle r^3 \cos(\phi - \psi_{1,3}) \rrangle $.

Our numerical results for the triangularity $v_3/\epsilon_3$ are similar to
recently reported  results \cite{Alver:2010dn,Schenke:2010rr}.
%(The results of \Refs{ } are compatible but not directly comparable). 
%These result show that $v_3/\epsilon_3$ is approximately equal $v_2/\epsilon_2$
%for sufficiently low freezeout temperature.  
However, $v_3$ (and $v_1$) is
significantly more sensitive to the freezeout dynamics.  To understand this we
studied the space time development of the triangularity (and dipole asymmetry)
in Figs.~\ref{epsilon_time} and ~\ref{flow_profile}.  These figures indicate
that the triangular flow develops on the same time scale as the elliptic flow.
(A similar conclusion for the triangular flow was reached in Fig. 3 of
\Ref{Alver:2010dn} based on kinetic theory calculations.) However, there is an
important difference between the elliptic flow and the dipole and triangular
flows which has not been fully clarified previously.  Specifically, the dipole
and triangular moments of the transverse flow grow quadratically with radius,
$u_{T}^{(3)}  \propto   r^2$, rather than linearly  as is the case with
elliptic flow, $u_{T}^{(2)} \propto r$. 
Consequently, edge effects can significantly reduce the dipole and triangular
flows.  
%In \Ref{ } the reduction of $\epsilon_n$ due to free streaming 
%for a period of  time was studied (see Fig. 9) , and higher $\epsilon_n$
%decreased more rapidly  at early times, and presumably this is simply 
%a rr
  Increasing the freezeout temperature cuts on the exterior region of the
flow profile, and therefore $v_{1}$ and $v_3$ are  more sensitive to the
precise freezeout criterion (see Figs.  ~\ref{vnvsnpart} and ~\ref{vnofpt}). 
This
unfortunate result may limit the usefulness of the dipole and triangular flows
in determining the shear viscosity of the quark gluon plasma.   Indeed the
strong reduction of the $v_3$ due to the shear viscosity
\cite{Alver:2010dn,Schenke:2010rr} is presumably largely due to the shear
viscosity below $T_c$, though this conclusion requires further investigation.

We also investigated the dipole asymmetry, $\llangle r^3 \cos(\phi - \psi_{1,3}) \rrangle$. 
The dipole asymmetry appears to the same order
in the gradient expansion  and has  not been studied previously to our knowledge.
The dipole asymmetry is   generally smaller than the triangularity 
since  $\epsilon_1$ is comparatively small.  However, $v_1/\epsilon_1$ 
is only marginally smaller than $v_2/\epsilon_2$ and $v_3/\epsilon_3$.
In non-central collisions the dipole asymmetry is strongly 
correlated with the triangularity   and the reaction plane 
as is illustrated in \Fig{correlation_fig}  and explained in 
\Fig{corrfig}.  We find that in non-central collisions the dipole asymmetry
is preferentially out of plane leading to a $v_1$ out of plane. The 
size of the observed correlation is somewhat smaller than the observed correlations
measured by the STAR collaboration and does not explain the charge asymmetry.

Finally, we noted that the strong correlation between the dipole asymmetry
and the triangularity can be measured experimentally by measuring two 
particle correlations with respect to reaction plane.  The final
result is a hydrodynamic prediction for a curious correlator 
\st
\label{CME2}
  \dlangle \cos(\phi_{\alpha} - 3 \phi_\beta + 2 \Psi_{PP})  \drangle  \, ,
\stp
which is shown in \Fig{coolplot}. This average is similar to averages
used to investigate the Chiral Magnetic Effect  (CME) and 
is no more difficult to  measure. 
The hydrodynamic prediction for  \Eq{CME2} 
is several  times larger than the correlation 
currently measured by the STAR collaboration, $\llangle \cos(\phi_{\alpha} + \phi_{\beta} - 2 \Psi_{PP})  \rrangle $. 
Thus, the proposed measurement is feasible
and important. If the predictions of \Fig{coolplot}  are confirmed it 
would validate the hydrodynamic and geometric nature of the measured two 
particle correlations.  Further, given the off-diagonal nature of  the proposed measurement, 
it will be difficult to reproduce this correlation with other mechanisms.

The current study neglected the effects of shear viscosity and  resonance
decays  and used an ideal gas rather than a lattice based equation of state. 
Incorporating  these important corrections is 
left for future work.

\section*{Acknowledgments} We gratefully acknowledge useful discussions
with Jean-Yves  Ollitrault, Sergei Voloshin, and especially Edward Shuryak. This
work is supported by an OJI grant from the Department of Energy
DE-FG-02-08ER4154 and the Sloan Foundation.

\appendix
\section{Details of the cumulant expansion and initial conditions }
\label{details}

\subsection{Formal expansion }
\label{cumulants_formal}

Our goal is to determine the cumulants of the underlying 
distribution $\rho(\x)$ and to decompose these cumulants into
irreducible tensors with respect to rotations around the $z$ axis.
 
First we expand $\rho(\x)$ and  its Fourier transform 
$\rho(\k)$ in a  fourier series
\begin{align}
 \rho(\x) = \rho(r, \phi) =& \rho_0(r) +  2\sum_{n=1} \rho_n^c(r) \cos(n\phi) + 
2 \sum_{n=1} \rho_n^s(r) \sin(n\phi) \, ,  \\
 \rho(\k) = \rho(k, \phi_k) =& \rho_0(k) +  2\sum_{n=1} \rho_n^c(k) \sin(n\phi_k)
 + 2 \sum_{n=1}^{\infty} \rho_n^s(k) \sin(n\phi_k) \, ,
\end{align}
where $r,\phi, k,\phi_k$ are the magnitudes  and azimuthal 
angles of $\x$ and $\k$ respectively.  The relation between
the $\rho_n^{c,s}(k)$ and $\rho_n^{c,s}(r)$ is established
by substituting the identity
\st
  e^{i\k\cdot \x} = J_0(kr) + 2 \sum_{n=1}^{\infty} i^n J_{n}(kr) \cos(\phi - \phi_k)  
\stp
into the Fourier transform (\Eq{fourier}) and using elementary manipulations 
to obtain
\st
  \rho_{n}^{c,s}(k)  = 2 \pi \int r \dd r\, i^n J_n(kr)  \rho_{n}^{c,s}(r) \, .
\stp

Similarly, the generating function of cumulants is  also given by  a fourier series
\st
 W(\k) =  W_0(k) + 2 \sum_{n} W_n^c(k) \cos(n\phi_k)  + 2 \sum_{n} W_n^s(k) \sin(n\phi_k) \, ,
\stp 
and each  $W_n^{c,s}(k)$ is expanded in $k$ as described by
equations Eqs.~\ref{Wnk_1} and \ref{Wnk_2}.  
Then we can expand  both sides of the defining relation
\st
   \exp(W(\k)) \equiv \rho(\k) \, ,
\stp 
in a series expressions of the form $k^m \cos(n \phi_k)$ and $k^{m} \sin(n\phi_k)$. In
developing this expansion we use the series expansion 
of the Bessel function  
\st
\mathop{J_{{\nu}}\/}\nolimits\!\left(z\right)=
(\tfrac{1}{2}z)^{\nu}\sum _{{k=0}}^{\infty}(-1)^{k}
\frac{(\tfrac{1}{4}z^{2})^{k}}{k!\mathop{\Gamma\/}\nolimits\!\left(\nu+k+1\right)}\, ,
\stp
and the series expansion of $W_{n}^{c,s}(k)$. 
Comparing idential powers of $k^m \cos(n\phi_k)$  and $k^{m} \sin(n \phi_k)$
we determine the $W_{n,m}^{c,s}$  in terms of the moments 
of the underlying distribution.  Through fifth order inclusive this 
comparison yields the following relations: \\

\noindent {\bf 0-th harmonic:}
\bg
  W_{0,2} &=& \frac{1}{2} \llangle r^2 \rrangle \,,  \\
  W_{0,4} &=& \frac{3}{8} \left[ \llangle r^4 \rrangle - 
2 \llangle r^2 \rrangle^2 - \underline{\llangle r^2\cos2\phi \rrangle^2} \right]   \, , 
\nd
{\bf 2nd harmonic:}
\bg
  W_{2,2}^c &=& \frac{1}{4} \left[ \llangle r^2 \cos2\phi \rrangle \right]   \, , \\
  W_{2,4}^c &=& \frac{1}{4} \left[ \llangle r^4 \cos2\phi \rrangle - 3 \llangle r^2 \rrangle \llangle r^2\cos2\phi \rrangle \right]  \, ,\\
  W_{2,4}^s &=& \frac{1}{4} \left[ \llangle r^4 \sin2\phi \rrangle \right]  \, ,
\nd 
{\bf 4th harmonic:}
\bg
   W_{4,4}^c = \frac{1}{16} \left[ \llangle r^4 \cos4\phi \rrangle -  \underline{ 3\llangle r^2\cos(2\phi) \rrangle^2} \right] \, , \\
   W_{4,4}^s = \frac{1}{16} \left[ \llangle r^4 \sin4\phi \rrangle  \right] \, ,
\nd
{\bf 1st harmonic:}
\begin{align}
  W_{1,3}^c =& \frac{3}{8} \left[ \llangle r^3 \cos(\phi) \rrangle \right] \, ,  \\
  W_{1,3}^s =& \frac{3}{8} \left[ \llangle r^3 \sin(\phi) \rrangle \right] \, , \\
  W_{1,5}^c =& \frac{5}{16} \Big[
 \llangle r^5 \cos(\phi) \rrangle 
- 6 \llangle r^2 \rrangle     \llangle r^3\cos\phi \rrangle \nonumber \\
  &  \qquad - \left(\underline{
\llangle r^2\cos2\phi\rrangle    \llangle r^3 \cos3\phi \rrangle 
+ 3 \llangle r^3\cos\phi \rrangle \llangle r^2\cos2\phi \rrangle 
} \right) \Big]  \, ,  \\
  W_{1,5}^s =& \frac{5}{16} \Big[ \llangle r^3 \sin(\phi) \rrangle  
- 
6\llangle r^2 \rrangle   
 \llangle r^3\sin\phi \rrangle    \nonumber \\
 & \qquad 
- 
\left(\underline{
\llangle r^2\cos2\phi \rrangle \llangle r^3 \sin 3\phi \rrangle  
- 3
\llangle r^3 \sin\phi \rrangle 
\llangle r^2\cos2\phi \rrangle 
} \right) \Big]  \, ,
\end{align}
{\bf 3rd harmonic:}
\bg
  W_{3,3}^c &=& \frac{1}{8} \left[ \llangle r^3 \cos(3\phi) \rrangle \right] \, ,  \\
  W_{3,3}^s &=& \frac{1}{8} \left[ \llangle r^3 \sin(3\phi) \rrangle \right] \,  , \\
 W_{3,5}^c &=& \frac{5}{32}\left[  \llangle r^5 \cos3\phi \rrangle - 4 \llangle r^2 \rrangle \llangle r^3 \cos3\phi \rrangle  - \underline {  6\llangle r^3\cos\phi \rrangle \llangle r^2\cos2\phi \rrangle }  \right]  \, , \\
 W_{3,5}^s &=& \frac{5}{32}\left[  \llangle r^5 \sin3\phi \rrangle - 4 \llangle r^2 \rrangle \llangle r^3 \sin3\phi \rrangle  - \underline {6 \llangle r^3\sin\phi \rrangle \llangle r^2\cos2\phi \rrangle }   \right] \, , 
\nd
{\bf 5th harmonic:}
\bg
 W_{5,5}^c  &=& 
\frac{1}{32} \left[ 
\llangle  r^5\cos(5\phi) \rrangle  
- \underline{10 \llangle r^2\cos2\phi \rrangle \llangle r^3\cos3\phi \rrangle }  \right] \, , \\ 
 W_{5,5}^s  &=& \frac{1}{32} \left[ 
\llangle  r^5\sin(5\phi) \rrangle  
- \underline{10 \llangle r^2\cos2\phi \rrangle \llangle r^3\sin3\phi  \rrangle }  \right] \, .  
\nd

Each coefficient has  a simple interpretation. For instance,
$W_{0,2} = \frac{1}{2} \llangle r^2 \rrangle$ 
is simply the root mean square radius of the Gaussian. To
classify corrections to the Gaussian, one should examine
the difference between $\llangle r^4 \rrangle $ and $\llangle  r^2 \rrangle^2$;
$W_{0,4}  \simeq
\frac{3}{8} \left[ \llangle r^4 \rrangle - 2 \llangle r^2 \rrangle^2  \right]   $ is the required  difference.
The underlined terms ({\it i.e.} $\underline{\llangle r^2\cos2\phi \rrangle^2}$ in the 
case $W_{0,4}$) are of suppressed by a power of $\epsilon^2$ and are therefore generally unimportant except in very peripheral collisions. 

%The precise relation is
%\st
%\left[ \frac{1}{W_{n,m}^{c,s}}  \frac{1}{p_T}  \frac{\dd \Delta N}{\dd y\,  \dd p_T\,  \dd(\phi_p -  \pi/(2n))}  \right]_{n,m,s} = 
%\left[ \frac{1}{W_{n,m}^{c,s}}  \frac{1}{p_T}  \frac{\dd \Delta N}{\dd y\,  \dd p_T\,  \dd(\phi_p -  \pi/(2n))}  \right]_{n,m,c} 
%\stp

\subsection{Fourier transform and regulating the cumulant expansion} 
\label{regulate}

After specifying the cumulants, the distribution is Fourier transformed 
to determine the initial entropy density in coordiante space.  For simplicity 
we will discuss only a spherically symmetric Gaussian deformed by 
a small definite triangularity, $W_{3,3}^c$.   In this case the 
Fourier transform of a distribution with $W_{0,2}$  and small 
$W_{3,3}^c$ , 
\st
    \rho(\x) = \int \frac{\dd^2\k}{(2\pi)^2} \, e^{-i \k \cdot \x} e^{-\frac{k^2}{2} W_{0,2} } \left[ 1  + \frac{1}{3!} W_{3,3}^c (ik)^3 \cos3 \phi_k + \ldots \right] \, ,
\stp 
yields with the definition of $\epsilon_3$ in \Eq{epsilon_3def} 
\st
\label{rho1}
   \rho(\x)  
= 
\left[1 +   \frac{\llangle r^3 \rrangle  \epsilon_3}{24} 
\left( \left(\frac{\partial}{\partial x} \right)^3 - 3 \left(\frac{\partial}{\partial y} \right)^2 \frac{\partial}{\partial x} \right) \right] \frac{ e^{-\frac{r^2}{\llangle r^2 \rrangle } } } {\pi \llangle r^2 \rrangle } \, , 
\stp
where $\llangle r^3 \rrangle = 3 \sqrt{\pi}/4 \, \llangle r^2 \rrangle^{3/2}$.
At large enough radius the correction term becomes large compared
to the leading Gaussian. To regulate this term we replace the whole correction ($\equiv X$) with 
\st
\label{rho2}
   X \rightarrow  C \tanh(X/C) \, , 
\stp
where $C = 0.95$. We have checked that the results are independent of 
the precise value of the constant $C$.  The regulator here is not 
perfect as it (weakly) mixes different terms in the fourier 
expansion, but we have found this to be unimportant from a practical perspective, i.e. the $v_2$ produced by this regulated $\epsilon_3$ distribution is
small.  Another complication is that the input parameter $\epsilon_3^{\rm input}$ in the regulated version of \Eq{rho1} does not actually equal the ``true" $\epsilon_{3}$ 
of the initial distribution. In all figures  we have divided by the true
$\epsilon_3\equiv -\llangle r^3 \cos(3\phi) \rrangle$ rather than
the input parameter $\epsilon_{3}^{\rm input}$.

We can now specify precisely the initial conditions that are used for \Fig{vnvsnpart} and other results.  
At a given impact parameter  we use the optical glauber model to 
calculate the distribution of participants the transverse plane  
with $\sigma_{NN} = 40\,{\rm mb}$.  In a traditional hydrodynamic simulation
(such labeled by the ``Glauber" curves in \Fig{cumulant_fig}) 
the entropy density at an initial time $\tau_o=1\,{\rm fm}$ is
\st
    s(x, y,\tau_0) = \frac{C_{s}}{\tau_0}\,\frac{\dd N_{p}}{\dd x\,\dd y},
\stp
where $\frac{\dd N_p}{\dd x\,\dd y}$ is the number of participants per
unit area.
The value  $C_{s}=15.9$ closely corresponds to the
results of full hydrodynamic simulations \cite{Teaney:2001av,Kolb:2000fh,Huovinen:2001cy} 
%and  corresponds to a maximum initial
%temperature of $T_{0} = 420\,\mbox{MeV}$ at impact
%parameter $b=0$. 
The equation of state that is used in this work 
is a classical massless ideal gas $\pr=1/3\, e$. The 
relation between the temperature and energy density is $e/T^4 \simeq 12.2$ which 
is the value for a two flavor ideal quark-gluon plasma. 
%In the 
%relevant temperature range $T\simeq 140 - 300\, {\rm MeV}$ 
%this is not too far from a lattice equation of state.
In the current simulations we calculate the total
entropy and average squared radius $\llangle r^2 \rrangle$ 
for glauber distribution. We then take a deformation $\epsilon_3 \simeq 0.1$,
and use  these parameters to initialize the regulated Gaussian 
described by \Eq{rho1} and \Eq{rho2}. Finally, the simulation is run and the spectra
are calculated leading to \Fig{envsnpart}.

\section{Correlations in the Glauber model}
\label{glauber}
The goal of this appendix is to motivate \Eq{p1p3pr}. 
A given distribution of participants is first characterized by the 
participant plane $\Psi_{PP} \equiv \psi_{2,2}$
and we will assume that $\epsilon_2$ is small. Then the probability
distribution for $\psi_{1,3}$ for fixed $\Psi_{PP}$ is given  by \Eq{Acoeffdef}.
For fixed $\Psi_{PP}$ and $\psi_{1,3}$ the probability distribution 
for $\psi_{3,3}$ must be $2\pi/3$ periodic. Measuring all 
angles with respect  to participant plane and keeping only the
first non-trivial term in the Fourier series we have
\st
\label{Papp} 
P(\psi_{3,3}|\psi_{1,3}\Psi_{PP})=\frac{1}{2\pi}\left[1+ 2B 
\cos\Big(3(\psi_{3,3}-\Psi_{PP}) - (\phi^*- \Psi_{PP}) \Big)\right] \, .
\stp
The amplitude $B$ and phase $\phi^{*}$ are functions of $\psi_{1,3} - \Psi_{PP}$.

\begin{figure}
\includegraphics[width=0.55\textwidth]{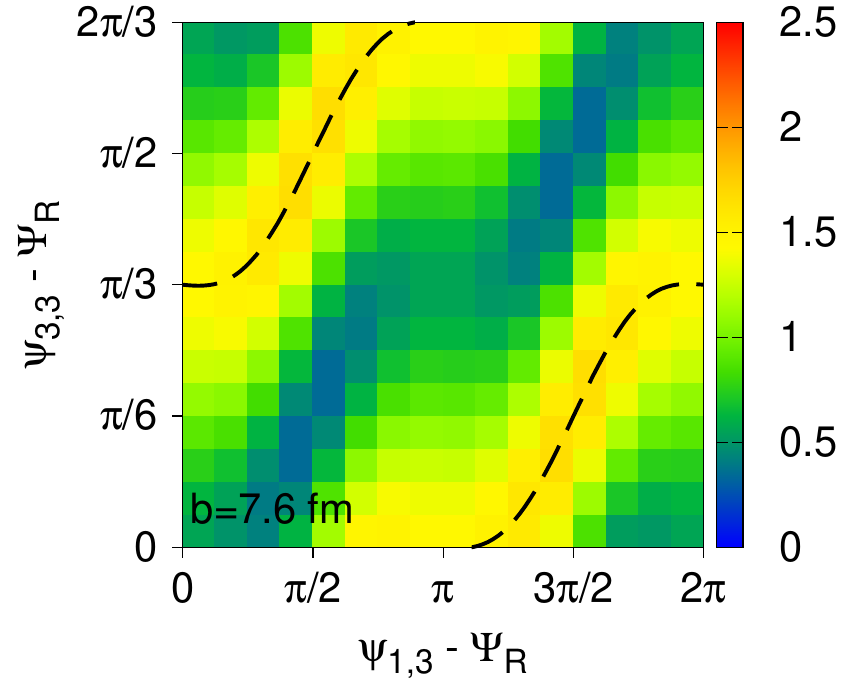}
\caption{A fit based on Eq.~\protect\ref{p1p3pr} to the 
the Glauber data exhibited in Fig.~\protect\ref{correlation_fig}. 
The parameters are $B_0 = 0.277(2)$, $B_2 = 0.029(1)$, and $C=0.532(7)$.
The normalization ({\it i.e.} the color scale) is arbitrary, but is the same as in Fig.~\protect \ref{correlation_fig}.
\label{fitplot}
}  
\end{figure}

The amplitude  $B$ and the phase derivative 
can be expanded in  a Fourier series
\begin{align}
\label{ampapp}
 B =& B_0 + 2 B_2\cos \left(2\psi_{1,3} - 2\Psi_{PP} \right)  \, ,  \\
\label{phasederiv}
\frac{\dd\phi^{*}}{\dd \psi_{1,3} } =& C_{0} + 2 C_2 \cos\left(2\psi_{1,3} - 2\Psi_{PP}\right)   \, .
\end{align}
As the $\psi_{1,3}$ increases by $2\pi$, the phase $\phi^{*}$ must change by a multiple of $2\pi$ to leave the conditional probability distribution invariant. The simplest possibility which qualitatively 
describes the trends
illustrated in \Fig{correlation_fig} and \Fig{corrfig} is to take $C_0 =1$. 
In a general fourier series of two variables other possibilities would 
be allowed, e.g. $C_0 = 3$. However such correlations turn out to be small
in the Glauber model.
Integrating
\Eq{phasederiv}  we find 
\st
\label{phaseapp}
 \phi = \psi_{1,3} +  C_{2} \sin(2\psi_{1,3} - 2\Psi_{PP} )  + \mbox{\rm const}
\stp
The constant required to reproduce \Fig{corrfig} is $\pi$.  The combination
of Eqs.~\ref{Papp}, ~\ref{ampapp}, and \ref{phaseapp} leads to the parameterization quoted in \Eq{p1p3pr}. In \Eq{p1p3pr} we absorbed the constant phase $\pi$ into the leading minus sign of $B_0$ and $B_2$ and changed the sign of $C_2$
so that all coefficients are positive in the final fit.
\Fig{fitplot} shows a fit to the Monte Carlo
Glauber shown in \Fig{corrfig} at $b=7.6\,{\rm fm}$ using this parameterization. The fit does
capture most of the essential features, but fails to reproduce
the sharpness of the correlation band.

Finally, we can
estimate the scaling of these coefficients with the average elliptic
eccentricity $\aveeps$.  
In a central collision $B(\psi_{1,3},\Psi_{PP})$ 
must vanish. This can be understood by examining \Fig{corrfig} and recognizing
that in a central collision there is no distinguishable difference between Position A
and Position B. The coefficient of $\cos(3\psi_{3,3} - \phi^* - \Psi_{PP})$  ({\it i.e.} $B$)
describes how  phase between the triangular and the dipole planes 
changes from  Position A to Position B. This coefficient  must vanish in central
collisions where Position A and Position B are identical.
Finally the coefficients $B_2$ and $C_2$ reflect the almond shape 
and must involve an additional power of $\aveeps$ relative to $C_0$ and $B_0$.  With these remarks we arrive at the scalings given in \Eq{epsscaling}. 

%%%%%%%%%%%%%%%%%%%%%%%%%%%%%%%%%%%%%%%%%%%%%%%%%%%%%%%%%%%%%%%%%%%%%%%%

\end{document}